\documentclass[longauth]{aa} 

\usepackage{graphicx}
\usepackage{txfonts}
\usepackage{amssymb}
\usepackage{afterpage}
\usepackage{placeins}
\usepackage{color}
\usepackage{url}
\usepackage{natbib}
\usepackage{multirow}
\usepackage[utf8]{inputenc}

\begin{document}
\title{Performance of the joint LST-1 and MAGIC observations evaluated with Crab Nebula data}
\titlerunning{Performance of LST-1 and MAGIC}


\author{\small
H.~Abe \inst{1} \and
K.~Abe \inst{2} \and
S.~Abe \inst{1} \and
V.~A.~Acciari \inst{3} \and
A.~Aguasca-Cabot \inst{4} \and
I.~Agudo \inst{5} \and
N.~Alvarez Crespo \inst{6} \and
T.~Aniello \inst{7} \and
S.~Ansoldi \inst{8,9} \and
L.~A.~Antonelli \inst{10}
% ...
}
\institute{
Institute for Cosmic Ray Research, University of Tokyo, 5-1-5, Kashiwa-no-ha, Kashiwa, Chiba 277-8582, Japan
\and Department of Physics, Tokai University, 4-1-1, Kita-Kaname, Hiratsuka, Kanagawa 259-1292, Japan
\and Instituto de Astrofísica de Canarias and Departamento de Astrofísica, Universidad de La Laguna, La Laguna, Tenerife, Spain
\and Departament de Física Quàntica i Astrofísica, Institut de Ciències del Cosmos, Universitat de Barcelona, IEEC-UB, Martí i Franquès, 1, 08028, Barcelona, Spain
\and Instituto de Astrofísica de Andalucía-CSIC, Glorieta de la Astronomía s/n, 18008, Granada, Spain
\and Grupo de Electronica, Universidad Complutense de Madrid, Av. Complutense s/n, 28040 Madrid, Spain
\and National Institute for Astrophysics (INAF), I-00136 Rome, Italy
\and INFN Sezione di Trieste and Università degli studi di Udine, via delle scienze 206, 33100 Udine, Italy
\and also at International Center for Relativistic Astrophysics (ICRA), Rome, Italy
\and INAF - Osservatorio Astronomico di Roma, Via di Frascati 33, 00040, Monteporzio Catone, Italy
% ...
}

   \date{Received ...; accepted ...}


\offprints{contact.magic@mpp.mpg.de, \mbox{lst-contact@cta-observatory.org}, \\$^{*}$Corresponding authors}

  \abstract
   {}
   {LST-1, the prototype of the Large-Sized Telescope for the upcoming Cherenkov Telescope Array Observatory, is concluding its commissioning in Observatorio del Roque de los Muchachos on the island of La Palma. 
The proximity of LST-1 (Large-Sized Telescope 1) to the two MAGIC (Major Atmospheric Gamma Imaging Cherenkov) telescopes permits observations of the same gamma-ray events with both systems. }
   {We describe the joint LST-1+MAGIC analysis pipeline and use simultaneous Crab Nebula observations and Monte Carlo simulations to assess the performance of the three-telescope system.
The addition of the LST-1 telescope allows the recovery of events in which one of the MAGIC images is too dim to survive analysis quality cuts. }
   {Thanks to the resulting increase in the collection area and stronger background rejection, we find a significant improvement in sensitivity, allowing the detection of 30\% weaker fluxes in the energy range between 200~GeV and 3~TeV. 
The spectrum of the Crab Nebula, reconstructed in the energy range $\sim$60~GeV to $\sim$10~TeV, is in agreement with previous measurements. 
}
   {}

   \keywords{Instrumentation: detectors --  Methods: data analysis --  Gamma rays: general    }

   \maketitle

\section{Introduction}
Very-high-energy (VHE, $\gtrsim 100$\,GeV) gamma rays cannot be observed directly in an efficient way due to their absorption by the atmosphere.
In turn, observations with space-born instrument are marred by relatively low fluxes at those energies. 
In the last three decades, imaging atmospheric Cherenkov telescopes (IACTs) proved to be sensitive instruments for the study of VHE gamma-ray emission from cosmic sources (see e.g., \citealp{2022Galax..10...21S} for a recent review). 
The combination of multiple telescopes at distances of the order of 100\,m (comparable to the size of the gamma-ray Cherenkov light pool), permits joint, stereoscopic analysis of the events, significantly improving the performance of the system \citep{1996APh.....5..119K}. 

The Cherenkov Telescope Array Observatory (CTAO) is the upcoming next-generation gamma-ray facility \citep{acha13} composed of two telescope arrays located in the Northern and Southern hemispheres. 
In order to cover a broad energy range (from few tens of GeV up to a few hundreds of TeV) it will be composed of telescopes of three different sizes: Large-Sized Telescopes (LSTs), Medium-Sized Telescopes (MSTs) and Small-Sized Telescopes (SSTs). 
The LSTs, with mirror diameters of 23\,m, will be the most sensitive part of the system for the lowest energy range of CTAO (tens of GeV). 
The construction of the first LST telescope, named LST-1, finished in October 2018. 
Since 2019 it is taking commissioning and engineering data \citep{2021JPhCS2156a2089F}.

LST-1 (Large-Sized Telescope 1) is located in Observatorio Roque de los Muchachos, La Palma (Spain), at the altitude of 2200 m a.s.l.. 
It is placed at a distance of only $\sim 100$\,m from the MAGIC (Major Atmospheric Gamma Imaging Cherenkov) telescopes, a pair of 17\,m diameter IACTs \citep{2016APh....72...61A}. 
Both systems work independently, but their proximity allows for an offline search of common events and enables joint LST-1+MAGIC analysis. 
A similar array containing telescopes of different sizes is being operated by the H.E.S.S. Collaboration \citep{2015arXiv150902902H}. 
However, in that case the difference in mirror area (approximately a factor of 5) causes a similar difference in the energy threshold.
On the other hand, in the case of LST-1+MAGIC combination, the difference between the mirror area of the LST-1 and one of the MAGIC telescopes is only a factor of 2.

In this work we report the common analysis chain of both instruments and its achieved performance using both Monte Carlo (MC) simulations and observations of Crab Nebula. 
In Section~\ref{sec:inst} we describe both participating instruments.
The used data and Monte Carlo (MC) simulations are described in Section~\ref{sec:data}.
We derive various performance parameters of the joint system and present them in Section~\ref{sec:perf}.
The concluding remarks are gathered in Section~\ref{sec:conc}.

\section{Instruments and data analysis}\label{sec:inst}
\begin{figure}[t]
    \centering
    \includegraphics[width = 0.49\textwidth]{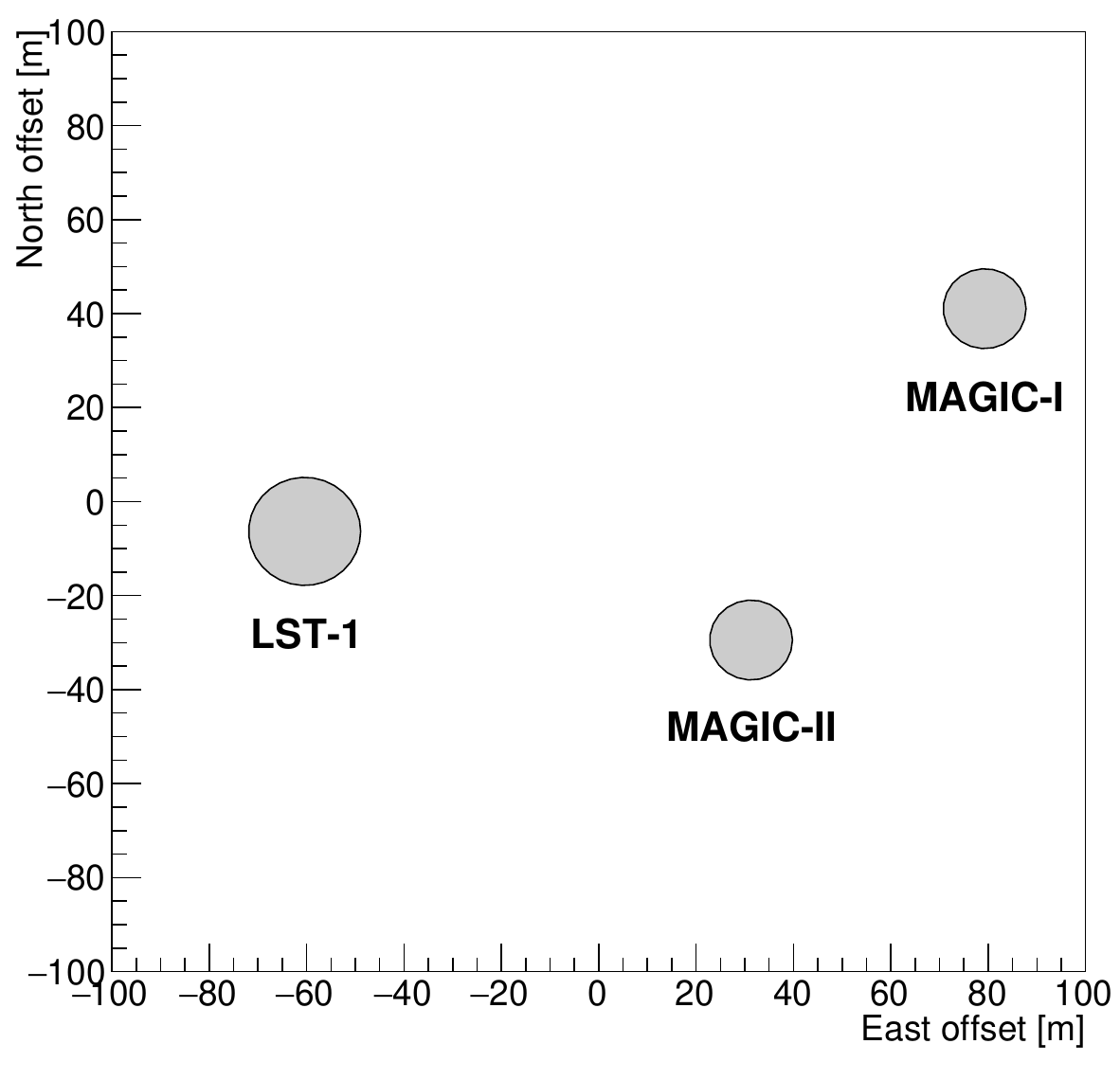}
    \caption{Location of the LST-1 and MAGIC telescopes. The X and Y axes represent the geographical East and North direction respectively. The diameter of the circle is equal to the diameter of the telescope's mirror dish.}
    \label{fig:tel_loc}
\end{figure}

The relative location of the LST-1 and the MAGIC telescopes, and their basic parameters are compared in Fig.~\ref{fig:tel_loc} and  Table~\ref{tab:pars}, respectively.
While the telescopes share the main design concepts, there are some differences, such as the larger LST-1 mirror area, its higher quantum efficiency (QE) of the optical detectors and its larger field of view (FoV). 
Despite the same parabolic dish shape (that minimizes the time spread of the registered Cherenkov photons) LST-1 has larger f/d  and larger camera FoV. 
The higher event rate in the case of LST-1 is a sum of multiple effects: lower threshold (due to higher QE and mirror area), larger size of the trigger region, and monoscopic operations. 
\begin{table}[t]
    \centering
    \begin{tabular}{c|c|c}
    Parameter     &  LST-1 & MAGIC I/II\\\hline
    Diameter (d) & 23\,m & 17\,m \\
    Focal length (f) & 28\,m & 17\,m \\
    Dish shape & parabolic & parabolic \\
    Camera FoV & $4.5^\circ$ & $3.5^\circ$ \\
    Pixel FoV & $0.1^\circ$ & $0.1^\circ$ \\
    Number of pixels & 1855 & 1039  \\
    Peak QE & 41\% & 32-34\% \\
    Sampling speed & 1\,GHz & 1.64\,GHz \\
    Trigger type & mono & stereo \\
    Typical event rate & $10^4$\,s$^{-1}$ & 300\,s$^{-1}$ \\
    Readout dead time & 7~$\mu$s & 26~$\mu$s \\
    \end{tabular}
    \caption{Comparison of LST-1 and MAGIC telescopes parameters.}
    \label{tab:pars}
\end{table}

\subsection{MAGIC}
MAGIC is a system of two IACTs, separated by a distance of 85\,m. 
The first telescope, MAGIC-I (M1), was constructed in 2003, and MAGIC-II (M2) was added in 2009.
Since then, both telescopes operate in a stereoscopic observation mode \citep{2012APh....35..435A}. 
In the standard operation mode only events triggering both telescopes are saved. 
The telescopes underwent a few upgrades, the most recent in 2012, and since then they share a nearly identical design and comparable performance. 
While the nominal camera FoV is $3.5^\circ$, the part covered by the trigger is limited only to the inner half of the full camera area. 
At low zenith distance, the energy threshold (defined as the peak of the differential true energy distribution) at trigger level of the MAGIC telescopes for a source with a $-2.6$ spectral index is $\sim 50$~GeV \citep{2016APh....72...76A} for a standard digital trigger.

\subsection{LST-1}

LST-1 is the first of the four LSTs to be constructed in the CTAO Northern site \citep{2022icrc.confE.872C}. 
The construction of LST-1 was completed in October 2018, after which its commissioning and validation period started. 
Currently the telescope is performing both commissioning and scientific observations. 
Nearly twice larger mirror area as well as improved QE of the optical sensors (photomultipliers) compared to MAGIC allow LST-1 to achieve an energy threshold of $\sim 20$~GeV \citep{lstperf}. 
However, as any IACT operating standalone, LST-1 suffers from huge hadronic background, which is much more efficiently rejected in stereoscopic systems.
Similarly, it also has worse accuracy of the reconstructed shower geometry, affecting angular and energy resolutions. 
Therefore, despite the larger light collection, at energies above 100~GeV, the sensitivity of LST-1 alone is a factor $\sim 1.5$ worse than that of MAGIC \citep{lstperf}. 

\subsection{Event matching}
Currently MAGIC and LST-1 operate independently. 
Both systems are however equipped with GPS clocks that provide time stamps for each event. 
Those time stamps can be used for offline matching of events that originate from the same shower (similar approach has been used in the first H.E.S.S. stereoscopic data, \citealp{2006NuPhS.151..373H}). 
Due to different electronic pathways and different travel times of the Cherenkov light to individual telescopes, a pointing-dependent time delay between the arrival times at MAGIC and at LST-1 needs to be taken into account.
For each subrun (corresponding to about 10~s of LST-1 data) we match the events with a coincidence window of 0.6~$\mu$s. 
The optimal delay is obtained using an iterative procedure. 
To allow also analysis of LST-1+M1 or LST-1+M2 event types, the procedure is done independently using time stamps in each of the MAGIC telescopes. 
For the typical rate of LST-1 and MAGIC (see Table~\ref{tab:pars}) this procedure would result in a negligible rate of accidental coincidences of $\lesssim 1.8$~s$^{-1}$. 
Anomalous coincidence combinations (such as matching two LST-1 events to one MAGIC event, or two MAGIC events to one LST-1 event) are excluded from the data stream, however they are very rare due to LST-1 and MAGIC  deadtimes. 

\subsection{Data analysis}\label{sec:data_analysis}
In their standalone operation both MAGIC and LST-1 are using independent analysis chains. 
The MAGIC data analysis is based on \texttt{MARS} \citep{2009arXiv0907.0943M, za13}, a C++, ROOT-based library and package of analysis programs.
The raw data are stored in a custom binary format, and data generated at each processing step are stored using ROOT containers. 

On the other hand LST-1 is using \texttt{cta-lstchain} \citep{lstchain}, a python-based analysis library exploiting \texttt{ctapipe} \citep{ctapipe}.
The LST-1 raw data consist of pixel-wise waveforms and auxiliary information. 
They are stored in a \texttt{zfits} format \citep{2012arXiv1201.1340P,2017ICRC...35..843L} and processed data are stored in  \texttt{HDF5} files \citep{2020SPIE11447E..0HN}.

\begin{figure*}[t!]
    \centering
    \includegraphics[width=0.99\textwidth]{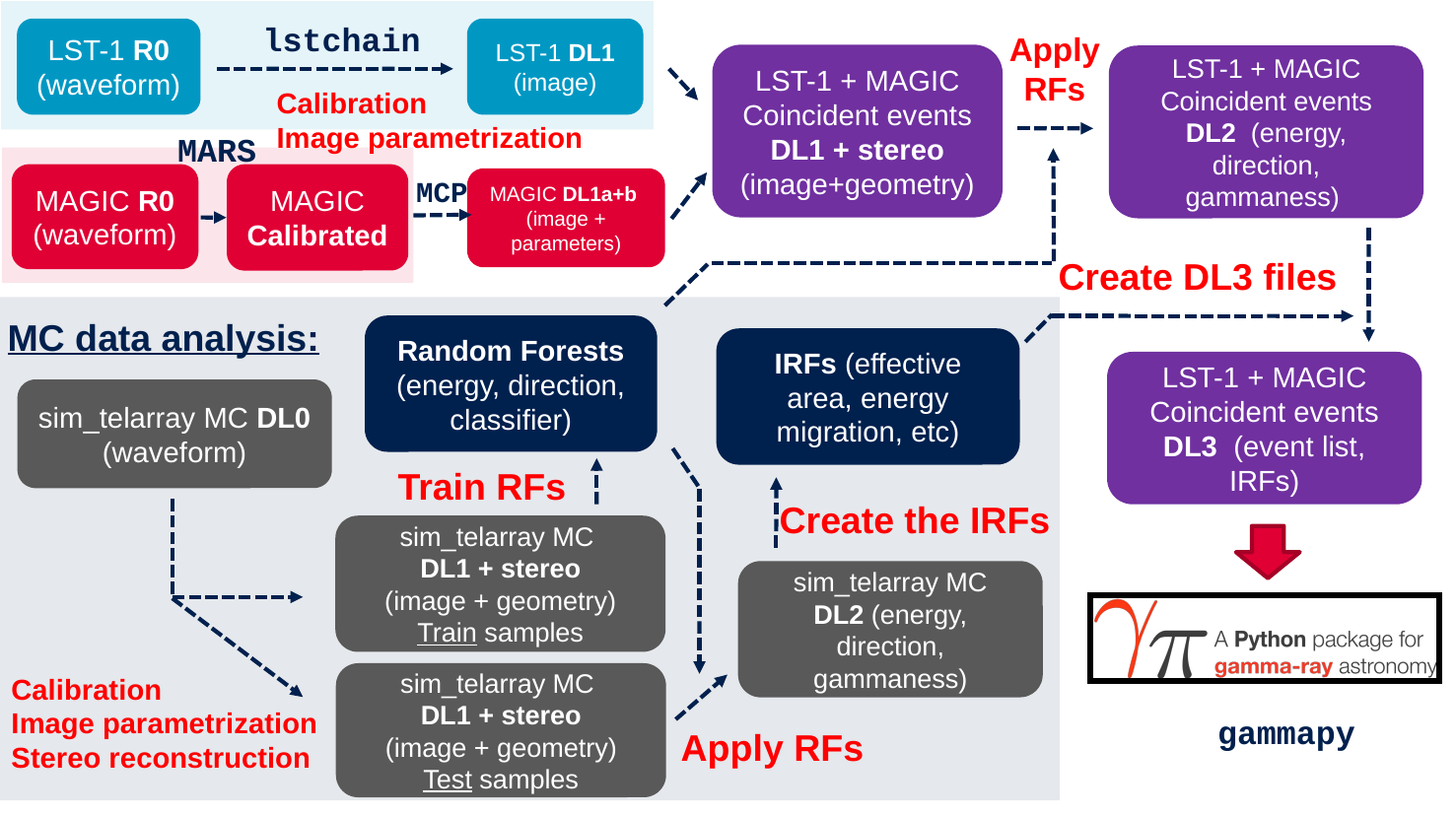}
    \caption{Schematic view of MCP analysis chain. Blue, red, violet and gray boxed represent different stages of LST-1 data,  MAGIC data, joint data and MC simulations respectively. 
    Black boxes mark the auxiliary files.   
    }
    \label{fig:chain}
\end{figure*}

For data obtained by observations, we perform the first stages of the data processing, namely the signal extraction from individual pixel waveforms and the calibration of the resulting images to photoelectrons (p.e.) and individual pixel timing, with the specific software of each instrument. 
Then the MAGIC data are converted into \texttt{HDF5} format, compatible with the LST-1 data using the dedicated \texttt{ctapipe\_io\_magic} package\footnote{\url{https://github.com/cta-observatory/ctapipe_io_magic}}.
The rest of the analysis chain is performed with the \texttt{magic-cta-pipe}\footnote{\url{https://github.com/cta-observatory/magic-cta-pipe}} package using \texttt{lstchain} and \texttt{ctapipe} methods. 
Namely, the \texttt{magic-cta-pipe} contains analysis scripts to, for example, apply the same image cleaning as in the MARS package, but within a \texttt{ctapipe}-like environment, and to match the events produced by the same shower in the three telescopes. All the other higher-level analysis steps are performed with the \texttt{magic-cta-pipe} package as well, with the help of other modules for specific tasks (e.g., \texttt{pyirf}, \citealp{2022icrc.confE.744N}, 
for the calculation of instrument response function, IRF and \texttt{gammapy} \citep{2017ICRC...35..766D} for flux estimation). For MC simulations, all the processing (including the calibration) is instead performed with \texttt{magic-cta-pipe}.
The analysis chain presented in this paper is referred to as the MAGIC-ctapipe (MCP) chain, and is summarized in Fig.~\ref{fig:chain}. 
Such a scheme allows us to profit from the automatic processing of the bulky, early stages of data and exploit the already implemented low-level calibration corrections (see e.g., \citealp{2013NIMPA.723..109S} for the case of MAGIC and \citealp{lstcalib} for LST-1). 
At the same time it permits the utilization of state-of-the-art software developed for CTAO, and in return, the newly developed tools can also be easily applied for CTAO analysis in the future.


After the initial image cleaning (see \citealp{2016APh....72...76A}) images are parametrized using the classical approach of \citep{1985ICRC....3..445H} and a quality cut on \textit{intensity} is applied (i.e., the total number of p.e. in the image should be at least 50 p.e.)\footnote{This is a standard cut both in MAGIC and LST-1 analysis chains, see \cite{2016APh....72...76A, lstperf}.}. 
Next, the events are divided into different classes, depending on which telescopes are triggered. 
Four combinations are considered: M1+M2, LST-1+M1, LST-1+M2, LST-1+M1+M2. 
However, it should be noted, that due to the stereoscopic trigger of the MAGIC telescopes, the event types LST-1+M1 and LST-1+M2 also correspond to events in which all three telescopes had been triggered.
Though in those events, one of the MAGIC telescopes provided an image that either did not survive the cleaning, or had too small intensity. 
In Table~\ref{tab:types} we report the percentage of events of each kind for various data and MC samples (the parameters of the MC simulations are given in Table~\ref{tab:mc_samples}). 
\begin{table}[t]
    \centering
    \begin{tabular}{c|c|c|c|c}
         Type &  MC $\gamma$ & MC $\gamma$  & MC p  & Observations \\
         & ($0.4^\circ$) & ($0-2.5^\circ$) & &  \\\hline
         M1+M2 & 6.2\% & 4.8\% &20.4\% & 21.5\% \\\hline
         LST-1+M1 & 7.1\% & 7.7\% & 6.2\% & 5.3\% \\\hline
         LST-1+M2 & 12.5\% & 12.6\% & 11.9\% & 14.2\% \\\hline
         LST-1+M1+M2 & 74.1\% & 74.8\% & 61.5\% & 59.0\% \\\hline
    \end{tabular}
    \caption{Percentage of different event types in different types of MC simulations and in the observations. \\
    \textit{Note:} Only images surviving 50 p.e. cut in \textit{intensity} are considered. Observations and MC simulations cover low zenith distance angle ($<30^\circ$). Proton MC are weighted to $-2.7$ spectral index, while gamma-ray MC to $-2.6$.
    Values for gamma-ray simulations are provided separated for showers at typical offset from the pointing direction ($0.4^\circ$) and for isotropic distribution (within $2.5^\circ$ from the pointing direction) }
    \label{tab:types}
\end{table}
\begin{table*}[]
    \centering
    \begin{tabular}{c|c|c|c|c|c|c|}
    Sample & Particle type & $Zd$ & $E_{\rm min}$ & $E_{\rm max}$ & Impact$_{\rm max}$ & Viewcone \\
    & & $[^\circ]$  & [GeV] & [TeV] & [m] & $[^\circ]$\\ \hline\hline
    \multirow{2}{*}{Train} & Gamma & 6 -- 52 & $5\times\cos^{-2.5} Zd$ & $50 \times\cos^{-2.5} Zd$ & $900\times\cos^{-0.5} Zd$ & 0 -- 2.5\\
    & Protons & 6 -- 52 & $10\times\cos^{-2.5} Zd$ & max($100 \times\cos^{-2.5} Zd$, 200) & $1500\times\cos^{-0.5} Zd$ & 0 -- 8 $\times\cos^{0.5} Zd$\\\hline  
    \multirow{4}{*}{Test} & Gamma & 10 -- 55 & $5\times\cos^{-2.5} Zd$ & $50 \times\cos^{-2.5} Zd$ & $700\times\cos^{-0.5} Zd$ & 0.4 \\
    & helium & 10 -- 43 & $20\times\cos^{-1.5} Zd$ & $200 \times\cos^{-1.5} Zd$ & $1500\times\cos^{-1} Zd$ & 0 -- 8 \\
    & Electrons & 10 -- 43 & $5\times\cos^{-2.5} Zd$ & $50 \times\cos^{-2.5} Zd$ & 720--1200 & $0 - 7.5$ \\
    \end{tabular}
    \caption{Summary of the generated MC samples and their zenith distance range, energy range, maximum impact parameter and viewcone (maximum offset angle from the camera center). \\
    \textit{Note:} The first three samples are the same used in \cite{lstperf}.}
    \label{tab:mc_samples}
\end{table*}
The dominating type of events are three-telescope events (3/4 of all gamma-ray events). 
About twice larger fraction of LST-1+M2 events than LST-1+M1 is related to the proximity of LST-1 to the M2 telescope. 
While the percentages of different event types in proton simulations roughly follow the one observed in the data, there are some minor differences at (absolute) 1--2\% level. 
They are likely caused by the presence of helium and higher elements in the data, as well as incompleteness of the simulations due to very large impact and offset angle events. 
Additionally, the regular systematic effects (light yield, optical points spread function, etc.) causing slight MC/data mismatches can also contribute to those small differences. 
The fraction of MAGIC-only (without the LST-1 counterpart) events is significantly larger in the observations and in the proton MC simulations ($\sim 20$\%) compared to the gamma-ray MC simulations ($\sim 5-6\%$, comparable for both point-like and diffuse gamma-ray simulations). 
We interpret this as a result of intrinsic differences between the Cherenkov light distribution on the ground for showers initiated by different primary particles.   
Gamma-ray-induced events have in general a smooth Cherenkov photon distribution on the ground. 
For such ``regular'' events, if they are bright enough to be detected by both MAGIC telescopes, the significantly higher light yield of LST-1 normally also allows the detection of the shower by the third telescope.  
However, hadronic events show irregularities in their ground distribution of Cherenkov light, caused by individual high transverse momentum sub-showers. 
Such events can produce a significant signal in MAGIC telescopes without an LST-1 event counterpart. 
Considering the small fraction of MAGIC-only events, and their dominant background origin, we exclude those events from further analysis. 

For convenience, and to exploit the information of telescopes not containing the image, the events are next divided into the combination types (see Table~\ref{tab:types}). 
For each event type and telescope participating in the combination, gamma/hadron separation parameter (\textit{gammaness}, see \citealp{lstperf}), estimated energy and the estimated  \textit{DISP} parameter (estimated distance of the source position projected on the camera to the centroid of the image, \citealp{2001APh....15....1L,2010A&A...524A..77A}) are computed.
The training is done using a Random Forest (RF) method \citep{2001MachL..45....5B}, implemented in the \texttt{scikit-learn} package \citep{2011JMLR...12.2825P}. 
The RF regressors used for energy and DISP estimations are using 150 estimators, a maximum tree depth of 50, the squared error criterion for selection of the best cut from all the parameters at each step and division of leaves down to a single event.
RF classifier used for gamma/hadron separation employed 100 estimators with a maximum depth of 100. 
In this case the RF branching is done using the Gini index criterion, but at each step, only the square root of the total number of parameters is randomly selected. 
Individual telescope estimates are based on the Hillas parametrization (\textit{intensity}, \textit{length}, \textit{width}, \textit{skewness}, \textit{kurtosis}, \textit{time slope} computed along the main axis of the image, and fraction of total image \textit{intensity} in the two outermost rings of pixels) in the particular telescope.
In each telescope, this information is combined with tentative stereoscopic parameters obtained from the axis crossing method \citep{1999APh....12..135H} (height of the shower maximum, impact parameter) and pointing direction (azimuth and zenith distance angles). 
In order to obtain event-wise classifiers and estimators, the individual telescope responses are weighted with the image \textit{intensity}\footnote{Other possible weights, including inverse of variance of the response of individual trees, were tested and proved comparable, but led to a slightly worse performance.}. 
In this way, brighter and better-reconstructed images are favored in the final estimation.

A special averaging procedure is applied for the estimation of the arrival direction of the shower. 
The arrival direction can be reconstructed from an image using the \textit{DISP} parameter, assuming that it lies on the main axis of the image in the camera plane. 
There are however two directions that fulfill this condition, located on opposite sides of the image. 
Selection of the correct one (the so-called head-tail discrimination), especially at the lower energies, may fail in a fraction of events. 
For example, \citealp{lstperf} reports that approximately 20\% of all gamma-ray events have head and tail wrongly discriminated for a spectrum similar to the Crab Nebula\footnote{This fraction is obtained after cleaning, intensity cut of 50 p.e. and with the main image axis oriented within $0.3^\circ$ of the nominal source position. It is however strongly dependent on energy, dropping below 5\% above 200~GeV.}.

Therefore, we apply the Stereo DISP RF method \citep{2016APh....72...76A}, adapted to three-telescopes observations. 
Namely, we scan all possible combinations of pairs of possible arrival directions from individual images and select the one that yields the smallest spread of reconstructed positions. 
The spread is quantified with the \textit{disp\_diff\_mean} parameter, defined as the sum of angular distances of reconstructed directions from all pairs of telescopes, divided by the number of such pairs.  
In order to enhance the angular resolution and provide additional rejection of hadronic events (which are more likely to have irregular images), we apply an additional cut of  \textit{disp\_diff\_mean}$<0.22^\circ$. 
The same value of the cut is used in the standard MARS analysis chain. 
The change of collection area at different stages of analysis (including application of the quality cuts) is summarized and discussed in Section~\ref{sec:Aeff}.

\subsection{Simulations of the telescopes' response to showers}
To train the shower reconstruction algorithm and to evaluate IRFs, the analysis of IACT data requires MC simulations.
In the case of LST-1, the development of showers is simulated using CORSIKA \citep{1998cmcc.book.....H}, while the response of the telescope is simulated with the \texttt{sim\_telarray} program \citep{2008APh....30..149B}.
On the other hand, within MAGIC, MC simulations of showers are generated using a slightly modified version of CORSIKA, but the response of the telescopes is obtained using \texttt{MagicSoft} programs (\texttt{reflector} and \texttt{camera}) \citep{2005ICRC....5..203M}.
Common LST-1+MAGIC observations require the analysis chain to be performed within the same framework, and the same should happen for the simulations.

We performed the simulations of the same showers visible by both MAGIC and LST-1, using the \texttt{sim\_telarray} program. 
To achieve this, we translated the simulation parameters of the MAGIC telescopes from the \texttt{reflector} and \texttt{camera} simulation programs into the \texttt{sim\_telarray} nomenclature.
For most of the parameters (e.g., mirror dish geometry, angular dependence of light guide efficiencies, average quantum efficiency of PMT, jitter of single p.e. times, telescope trigger parameters, readout pulse shape), the translation was direct and the same values/curves were used in both simulation chains. 
For some of the parameters, however, minor simplifications or averaging had to be applied due to intrinsic differences between the two softwares.  For example, this was the case for the simulation of the mirrors reflectivity and the noise within the pulse integration window.
Thanks to the usage of \texttt{sim\_telarray}, the LST-1 simulation parameters could be taken directly from the standard configuration, the so-called LSTProd2 (as in \citealp{lstperf}). 
The common parameters (atmospheric model, geomagnetic field) follow the LSTProd2 settings. 
The level of uniform night sky background (NSB) was adjusted at the analysis level in the case of LST-1, following the procedure described in \cite{lstperf}. 
In the case of MAGIC, the adjustment was done according to the same principle (matching noise in empty, the so-called pedestal, events), but already at the telescope camera response simulation level.
The validation process of the MC simulation settings on a dedicated MC production is described in Section~\ref{sec:mccomp}.
As a final end-to-end check we also compared the energies of MAGIC events reconstructed with both the MCP (based on the \texttt{sim\_telarray} MCs) and MARS (based on standard MAGIC MCs) chains, achieving similar accuracy (see \ref{sec:Monly}).

\section{Observation and simulation samples}\label{sec:data}

We determine the performance of the joint analysis chain in two ways: using observation data taken from the direction of the Crab Nebula and using dedicated MC simulations, respectively.

\subsection{Observations}
In order to evaluate the performance of the joint analysis, we used 4\,hrs of good quality Crab Nebula data taken simultaneously by the LST-1 and MAGIC telescopes.
The observations span the period between October 2020 and March 2021, and were taken in wobble observation mode, with the source position offset by $0.4^\circ$ from the camera center. 
After every 20-min long run, the direction of the offset was flipped to maintain consistency between the source and the background control region. 
Only data in which the pointing direction of both systems matched within $0.1^\circ$ 
were used. 
The data was taken at low and medium zenith angles, namely 
0.8\,hrs in between $12^{\circ} - 30^{\circ}$, 
2.3\,hrs in  $30^{\circ} - 45^{\circ}$ 
and 0.6\,hrs in between $45^{\circ} - 53^{\circ}$. 

\subsection{MC simulations}

For most of the analysis chain we re-use the same MC simulation samples of protons and gamma rays as in \cite{lstperf}.
However, we also used additional simulations of helium and electrons, with reoptimized scaling of the simulation parameters (maximum impact parameter and offset angle from the camera center) with zenith angle distance, to improve the sample completeness. 
The samples were generated at fixed pointings along the path of the source in the sky (training samples), or to cover the full-sky on a grid of pointings (test samples), see \cite{lstperf} for details. 
All the MC samples are generated with spectral index of $-2$ and reweighted to specific particle spectra. 
The productions are summarized in Table~\ref{tab:mc_samples}.
%
In the interest of studying the performance with MC as well, we have divided the ``TrainProton'' sample 
into training and testing sub-samples.

\subsection{Data/MC comparisons}
To ensure the correct reproduction of observed data by the MC simulations, we perform an end-to-end comparison with the data. 
As the gamma-ray showers are more regular and on average less extended than hadronic ones, such comparisons performed with gamma-ray events are sensitive to possible data/MC mismatches. 
Hence, we present a comparison with selected gamma-ray events, which also reflect the performance for gamma-ray observations. 
Nevertheless, for completeness of the study and to validate the analysis threshold, we also perform similar comparisons with the background events (see \ref{sec:comp_bgd}).

We derive the parameter distributions obtained from the gamma-ray excess events. 
The distributions are extracted from Crab Nebula observations after subtraction of the residual background using a background control region.
We compare these excess distributions to the simulated gamma rays weighted (and normalized) according to the spectrum that was measured by \cite{2015JHEAp...5...30A}.
In this approach, the gamma-ray events are dominated by a background of much more abundant hadronic showers.
Thus, to avoid large statistical (and systematic) errors some kind of background suppression needs to be applied. 
In order to perform the comparison without introducing a large bias, we applied soft cuts corresponding to 95\% \textit{gammaness} efficiency (in each estimated energy bin), and considered only events with reconstructed direction up to $0.2^\circ$ away from the nominal source position.

\begin{figure*}[pt!]
    \centering
    \includegraphics[width = 0.99\textwidth]{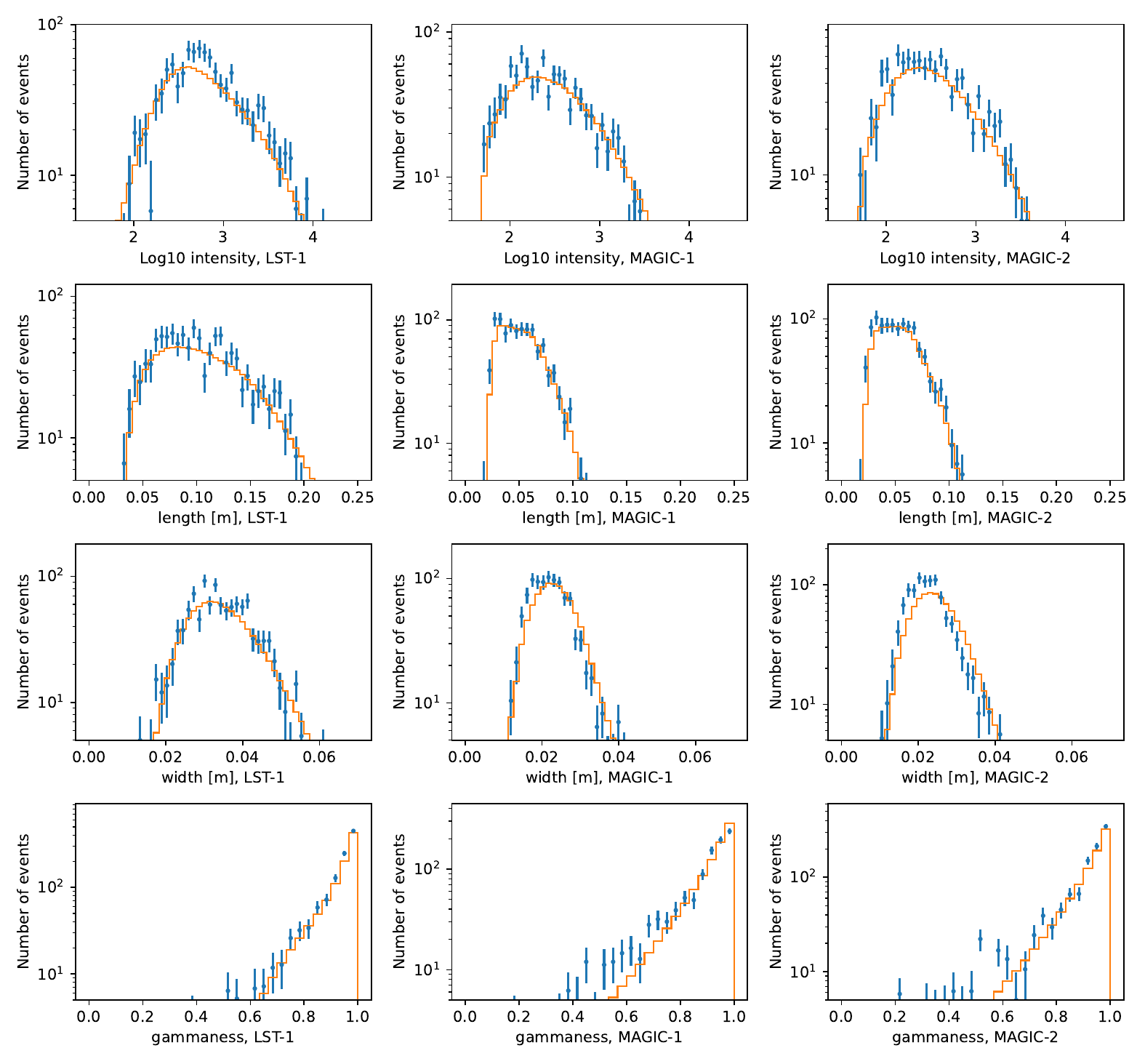}
    \includegraphics[width = 0.99\textwidth]{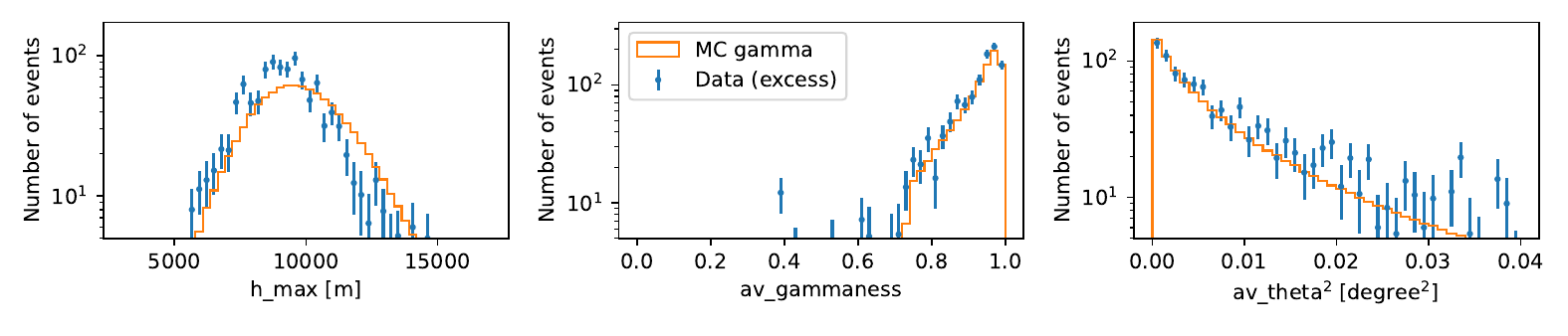}
    \caption{Comparison of image parameters between the gamma-ray excess in the data (blue) and MC simulations of gamma rays (orange). 
    Only observations with zenith distance below 30$^\circ$ are used.
    The top four rows of panels show \textit{intensity}, \textit{length}, \textit{width} and individual telescope \textit{gammaness} (from top to bottom) for LST-1 (left), MAGIC-I (middle) and MAGIC-II (right).
    The bottom row shows stereoscopic parameters: height of the shower maximum (left), averaged \textit{gammaness} (middle) and squared reconstructed distance to the source (right). }
    \label{fig:comp_gamma}
\end{figure*}

The results of the comparison are shown in Fig.~\ref{fig:comp_gamma}.
The \textit{intensity} distribution is roughly reproduced. 
We note that the MC simulation shape of the distribution of these parameters (in particular \textit{intensity}) is dependent on the assumed spectral model of the Crab Nebula. 
The \textit{length} distribution is well matching between the data and MCs. 
However, contrary to the background case (c.f.~\ref{sec:comp_bgd}), the \textit{width} parameter is slightly underestimated in the case of MAGIC-I and MAGIC-II telescopes, that could be e.g., due to insufficiently accurate simulation of the optical PSF.
Despite this, the \textit{gammaness} distribution, both for individual telescopes and average, is still sufficiently well reproduced in the simulations to avoid introducing large systematic errors. 
The reconstructed height of the shower maximum is slightly shifted towards higher values in MC simulations than in the data.
This could be due to a combination of various effects, e.g. systematic uncertainty of the energy scale of the telescopes, mismatch between the zenith/azimuth distributions which are continuous for the data and discrete for the simulations, slight mispointing of the telescopes. 
Finally, while the reconstruction of the event direction is roughly consistent with the simulations, a slight increase of the high-values tail in the data is present as well. 
Similarly, a slight mismatch in such distributions has been observed in LST-1-alone and MAGIC-alone observations, and might be related to arcmin-scale mispointing of the telescopes \citep{2016APh....72...76A,lstperf}. 

\section{Performance parameters}\label{sec:perf}

Using Crab Nebula data and MC simulations, we evaluate various performance parameters of the joint analysis chain and compare it with the MAGIC-only analysis. 
We also compare the sensitivity and flux reconstruction with the LST-1-only analysis. 
As the performance of Cherenkov telescopes is strongly dependent on the zenith distance of the pointing, we investigate separately the case of $Zd<30^\circ$ and $30^\circ<Zd<45^\circ$, and provide a comparison with the MAGIC-only performance.


\subsection{Energy threshold}
\begin{figure}[t]
    \centering
    \includegraphics[width=0.49\textwidth]{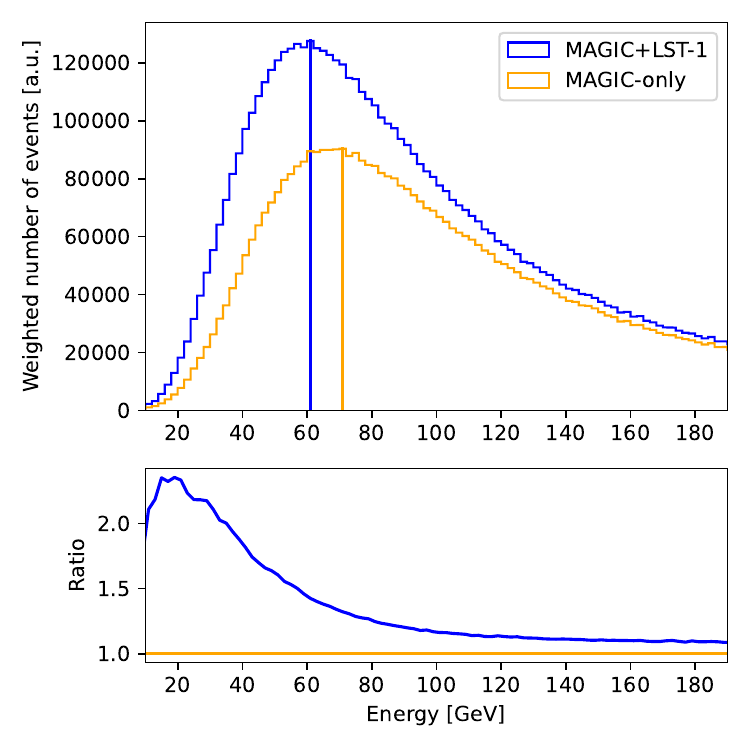}
    \caption{True energy distribution obtained with MC simulations (weighted to a source spectrum of $-2.6$) of gamma rays for $Zd<30^\circ$ at the reconstruction level (at least two images with \textit{intensity} $>50$). Vertical lines show the peak position for the joint analysis (blue) and MAGIC-only analysis (orange). 
    Bottom panel shows the ratio of the two curves.}
    \label{fig:threshold}
\end{figure}

In Fig.~\ref{fig:threshold} we present the differential true energy distribution for a source with a $-2.6$ spectrum.
In the case of MAGIC-only events, the energy threshold (peak position of that distribution) at the stereoscopic reconstruction level of $\sim 70$~GeV is consistent with the value obtained in \cite{2016APh....72...76A}.
The addition of the third telescope, while it cannot provide additional events at the trigger level, can recover events in which one of the MAGIC images has too small \textit{intensity} for further stereoscopic reconstruction. 
As a result, the energy threshold at reconstruction level is reduced to $\sim 60$~GeV.
Additionally, the collection area below the energy threshold is greatly improved, by a factor of about 2 at 30~GeV. 

\subsection{Flux reconstruction}
\begin{figure}[t]
    \centering
    \includegraphics[width=0.49\textwidth]{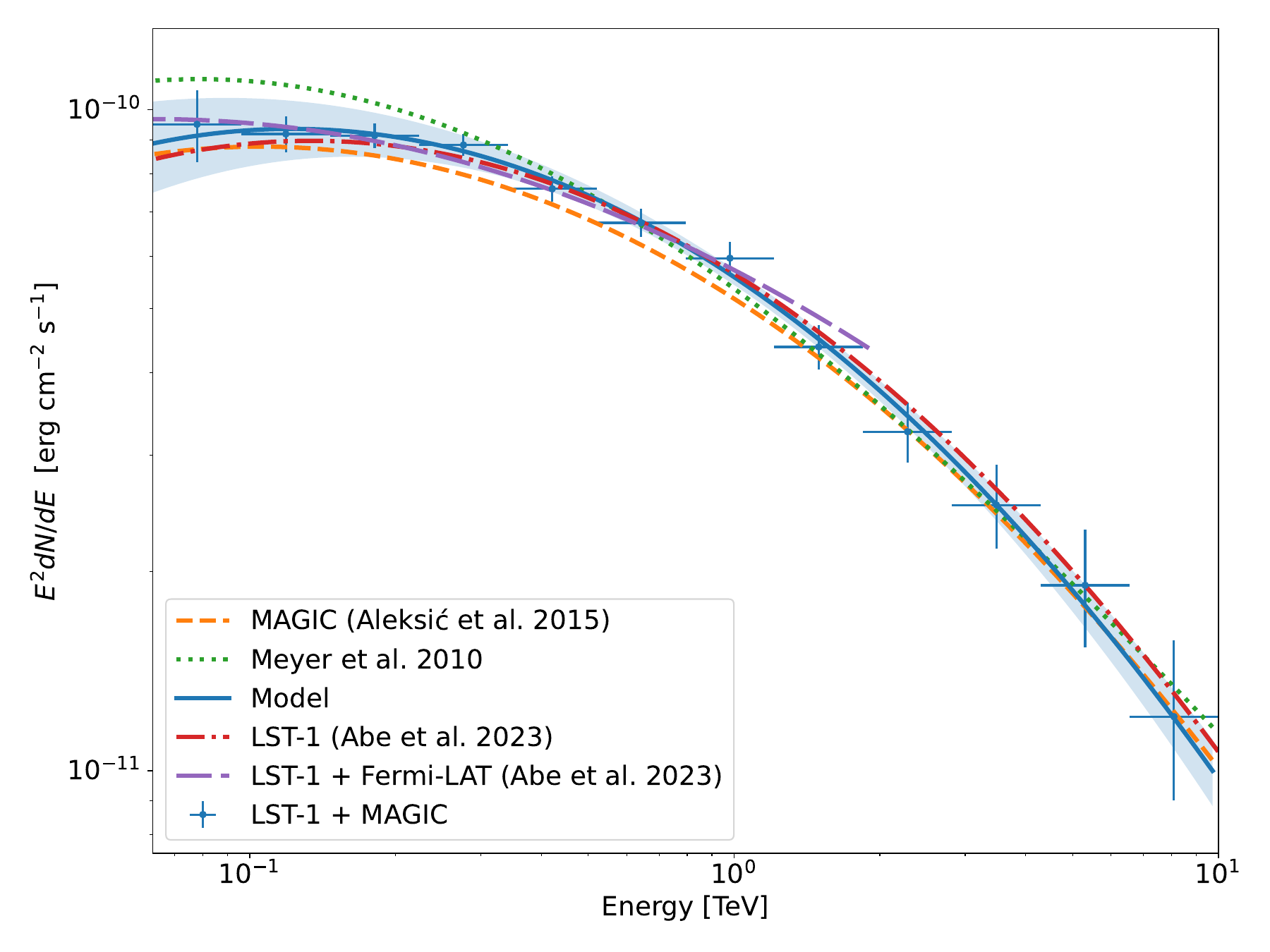}
    \includegraphics[width=0.49\textwidth]{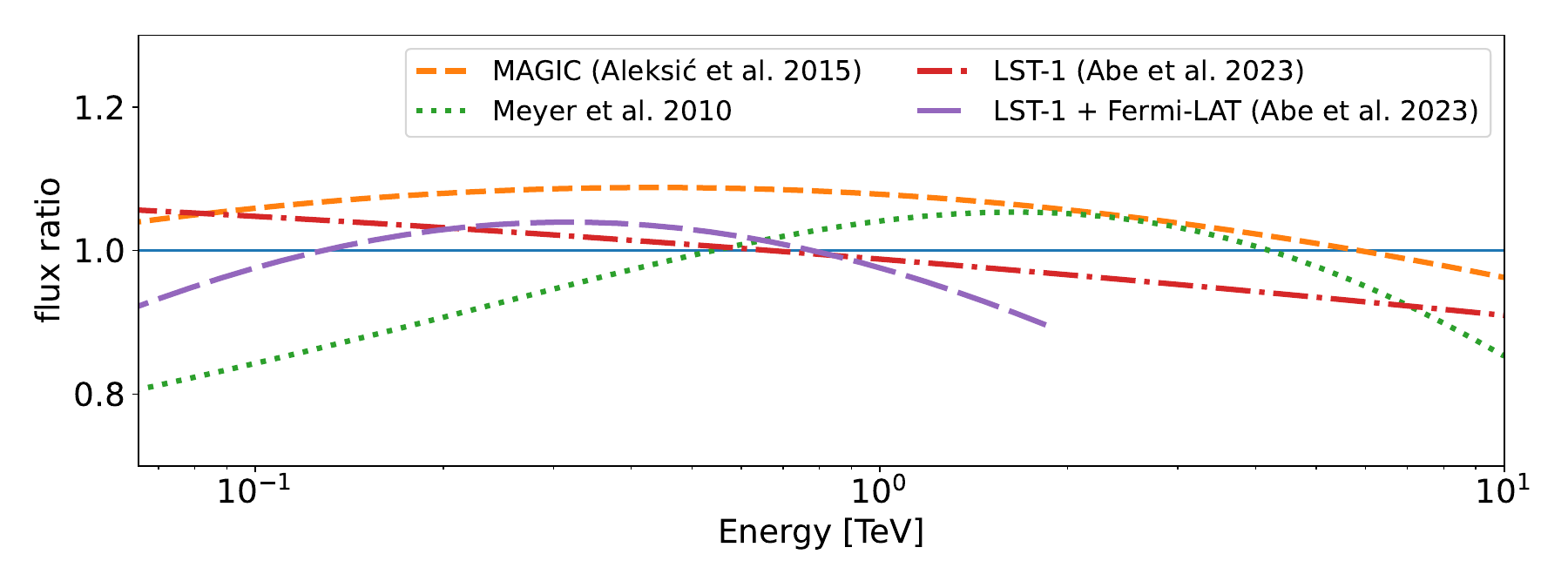}
    \caption{Spectral energy distribution of Crab Nebula obtained with joint LST-1+MAGIC analysis (blue, points and fit line, with the statistical uncertainty of the fit shown as shaded region) compared to reference measurements from 
    MAGIC-alone (orange dashed line, \citealp{2015JHEAp...5...30A}), 
    LST-1-alone (red dot-dashed, \citealp{lstperf}),
    \textit{Fermi}-LAT+LST-1 (violet long-dashed, \citealp{lstperf}),
    and  \textit{Fermi}-LAT+IACT (green dotted, \citealp{2010A&A...523A...2M}).  
    The bottom panel shows the ratio of the spectral model derived with the joint analysis to the individual reference spectra (see the legend). 
    }
    \label{fig:sed}
\end{figure}

Since all the data used in this work were taken before August 2021, following \cite{lstperf} for the spectral analysis we apply an increased cut of $intensity > 80$~p.e. for LST-1 images.
Due to the significantly larger light yield of LST-1 compared to MAGIC, the effect of this cut on the stereoscopic analysis is very small (e.g., for low zenith angle observations at 30~GeV only 10\% of events are removed). 
For MAGIC images, a standard $intensity > 50$~p.e quality cut is applied.
To reconstruct the spectrum of the Crab Nebula, we derive the IRFs for a number of simulated azimuth/zenith pointings close to those followed by the source during the observations.
To evaluate the IRFs corresponding to individual data runs we employ interpolation. 
We then divide the sample into an ascending and descending branch (i.e., before and after the culmination). 
For each branch separately, we perform a one-dimensional interpolation (over the cosine of zenith distance angle) of the IRFs. 
Subsequently a global, binned, joint likelihood spectral fit is performed with \texttt{gammapy} 0.20.1 \citep{2017ICRC...35..766D,gammapy20} to determine the best parameters of the spectral model for a point-like source at the nominal position of the Crab Nebula.
Next, the same software is used to derive individual spectral points by fitting the normalization of the global model in energy bands.  
In Fig.~\ref{fig:sed} we present the resulting spectral energy distribution reconstructed between $\sim$60~GeV and $\sim$10~TeV from the total investigated data set.
The spectrum is modeled in \texttt{gammapy} with a log parabola spectrum defined as:
\begin{equation}
    dN/dE=A (E/E_0)^{-\alpha - \beta \ln (E/E_0)}, 
\end{equation}
with $A=(3.48\pm0.09_{\rm stat})\times 10^{-11}\mathrm{ cm^{-2}s^{-1} TeV^{-1}}$, 
$E_0=1$~TeV, $\alpha=2.49\pm0.03_{\rm stat}$, $\beta=0.117\pm0.017_{\rm stat}$
\footnote{Note that in \cite{2016APh....72...76A} the log parabola is defined using the logarithm with base 10, which explain the very different values reported for the $\beta$ parameter.}.
The resulting spectrum is consistent with previous MAGIC and LST-1 measurements within $\sim 10$\%.

\begin{figure}[t]
    \centering
    \includegraphics[width=0.49\textwidth]{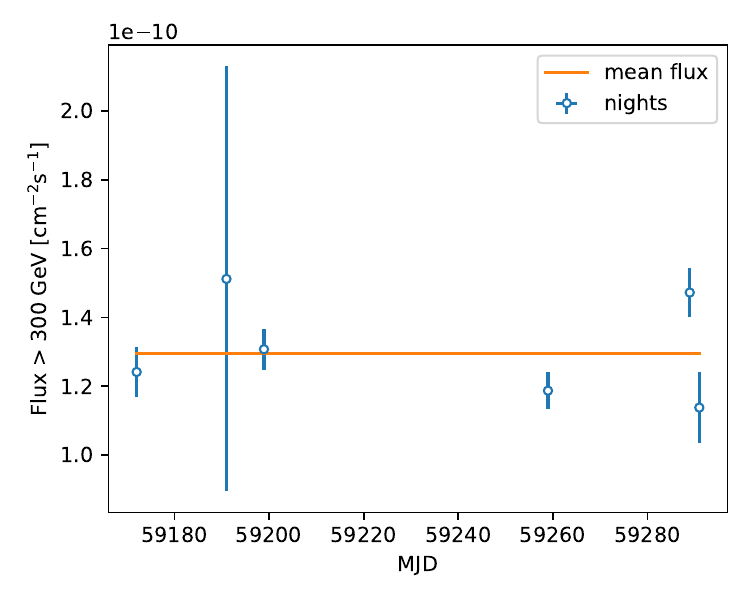}
    \caption{Integral flux of Crab Nebula obtained with joint LST-1+MAGIC analysis binned day-by-day (blue empty points). The horizontal line shows the corresponding average flux from the integrated spectral model.     
    }
    \label{fig:lc}
\end{figure}

In order to evaluate the stability of the flux reconstruction, we compute a light curve of the observed flux above 300~GeV (see Fig.~\ref{fig:lc}).
The plotted data are binned night-by-night, however we also investigated the stability of the flux at the run-by-run (corresponding to $\leq$20~min per bin) time scales. 
Similarly to other IACT measurements \citep{2006A&A...457..899A,2016APh....72...76A,lstperf}, the resulting Crab Nebula light curve is not consistent with a constant fit ($\chi^2/N_{\rm dof}=39.5/15$ for the run-by-run calculations and 13.1/5 for night-by-night).
Such observed instability of the flux is likely due to the systematic effects related to e.g., the atmosphere varying during the observations. 
We investigated how the $\chi^2/N_{\rm dof}$ statistics changes when a given level of systematic uncertainties is added in quadrature to the statistical uncertainty. 
To achieve the corresponding fit probability of 0.5, an additional 12.7\% systematic uncertainty is required in the case of run-by-run analysis and 7.9\% in the case of night-by-night, which is at the level or even lower than what was estimated for MAGIC.

\subsection{Differential flux sensitivity}\label{sec:sens}

Sensitivity is a measure of the minimum flux of a source that can be detected with an instrument in a given time exposure. 
In the case of differential sensitivity the detection should be achieved independently in a particular energy bin. 
We follow the definition typically used in the IACT community, namely, the data are divided into 5 bins per decade of reconstructed energy and the event statistics are rescaled to 50\,hrs of observation time. 
Sensitivity in a given bin then corresponds to the gamma-ray flux from a point-like source that provides $5\sigma$ significance signal, as computed via the equation 17 of \citet{1983ApJ...272..317L}, with additional two conditions: the number of excess events should be larger than 10 and also larger than 5\% of the residual background. 
In the calculations of the significance we assume that the background can be computed from 5 regions with the same acceptance of the signal region.

In order to optimize the usage of statistics, we apply a k-fold cross-validation procedure (see e.g., \citealp{mt68}). 
Namely, the sample is divided into four sub-samples, each of them using every fourth event in the sample. 
For each of them, we apply cuts in arrival direction and gammaness, computed using the remaining sub-samples and optimized to provide the best sensitivity. 
We then stack the events from all sub-samples and compute the final sensitivity that is not biased by the cut selection.

We estimate the sensitivity both with the Crab Nebula data and also with MC simulations.
In the latter case we use the spectrum derived by \cite{2015JHEAp...5...30A} to convert the flux into Crab Nebula units (C.U.). 
In the calculations we assume the proton flux of \cite{2019ICRC...36..163Y} and the electron flux is a parametrized combination of \textit{Fermi}-LAT and H.E.S.S. all-electron spectrum applying the parametrization of Equation 2 in \cite{2021JPhG...48g5201O}.

We follow the approach of \cite{2012APh....35..435A} to include the effect of the other elements. 
Namely we use helium simulations and scale the spectrum to 0.8 of the proton spectrum to also roughly take into account the heavier elements. 
It should be noted, however, that helium and higher elements have very little effect on the sensitivity. 
As it can be seen in Fig.~\ref{fig:comp_bgd}, at high \textit{gammaness} values, their contribution to the estimated background is smaller by about an order of magnitude than the one of protons (see also  \citealp{2018APh....97....1S}). 
It also drops very fast with estimated energy. 
High rejection of proton events via the \textit{gammaness} cut at high energies results in severely reduced background statistics.
Therefore, we collect the background statistics from the inner $1^\circ$ radius region and apply an energy-dependent correction factor between average proton density in this region and the density at camera offset of $0.4^\circ$.  
The correction factor is computed using a loose (corresponding to 94\% efficiency for gamma rays) \textit{gammaness} cut and is typically $\sim 1.4$.

When calculating the sensitivity using the Crab Nebula data, the background is taken around a direction in the sky with the same angular offset from the camera center as the source.  
For energies below 400~GeV, where background is abundant, we use only the reflected source position to minimize systematic uncertainties, while above this value we use 5 background estimation regions.
This approach also protects against overlapping background estimation regions at the lowest energies. 
Moreover, above 600~GeV, where the background is very scarce, and high angular resolution make the optimal angular distance cut from the source small, we increase the background statistics by using a broad cut for the background estimation region ($\theta<0.2^\circ$), and scale the number of events to the actual $\theta$-cut.

\begin{figure*}[t]
    \centering
    \includegraphics[width=0.49\textwidth]{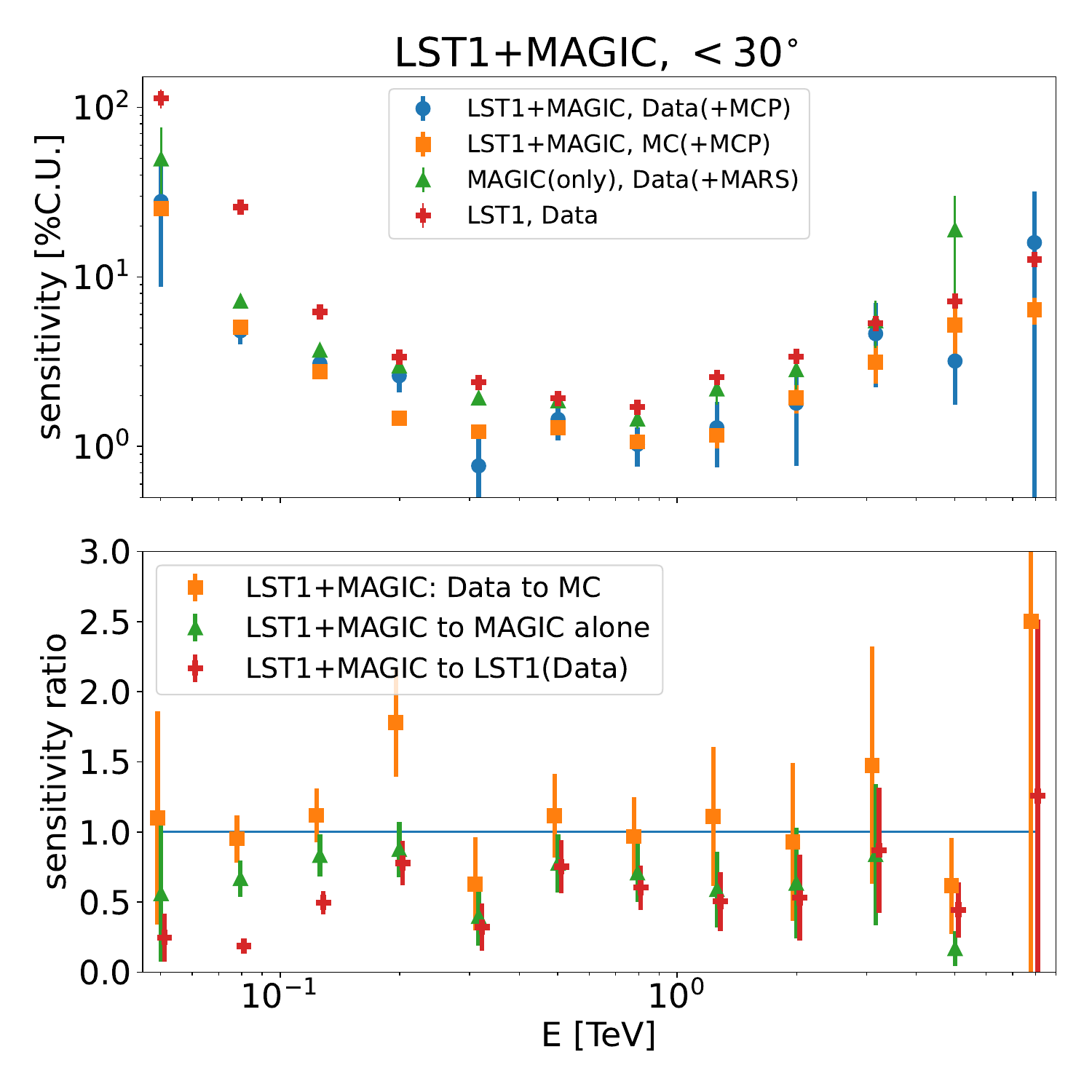}
    \includegraphics[width=0.49\textwidth]{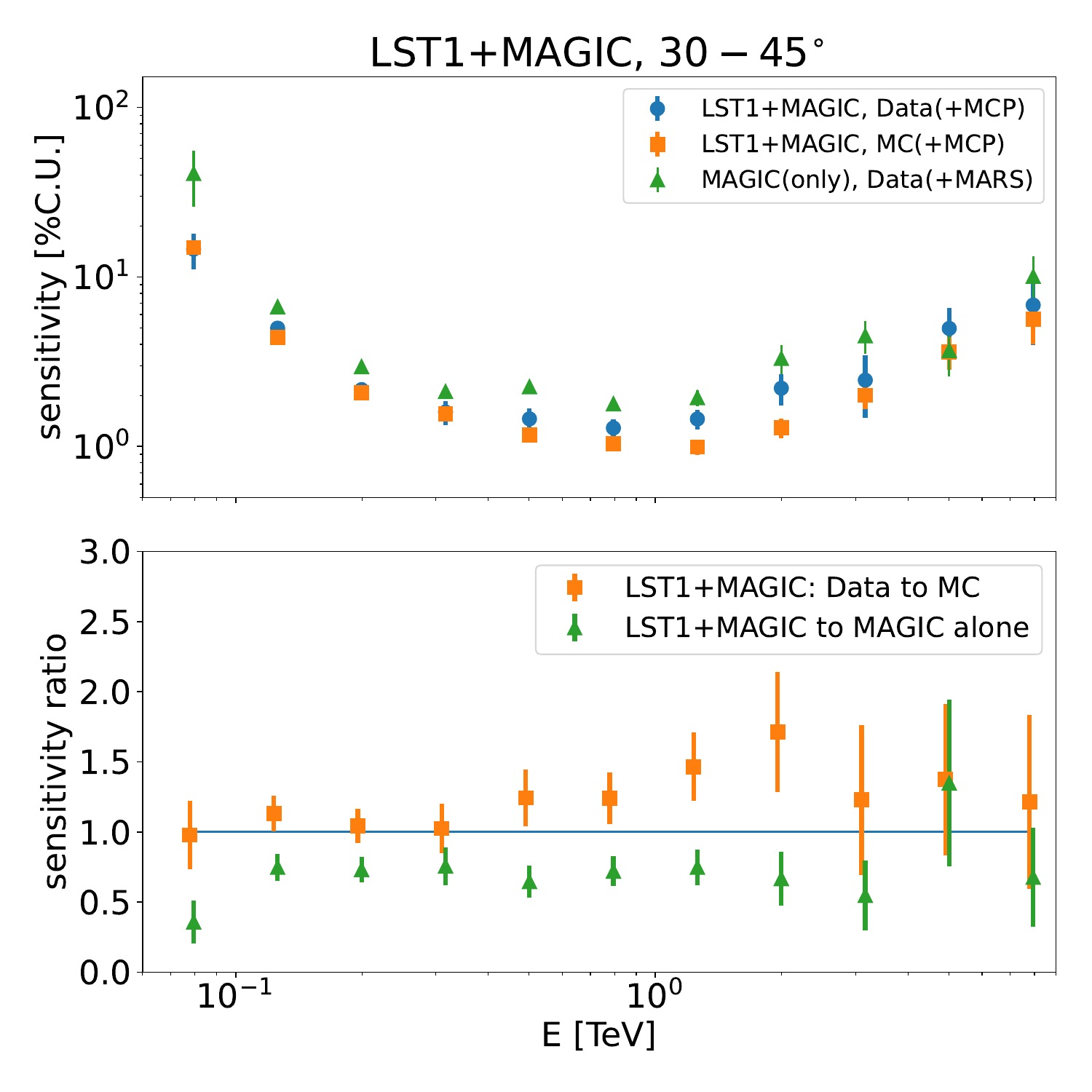}
    \caption{Sensitivity in the units of Crab Nebula flux percentage of the joint LST-1+MAGIC analysis with MCP from the Crab observations (blue circles) and MC simulations (orange squares) compared to MAGIC-only analysis using MARS (green triangles) and LST-1 sensitivity (red crosses, \citealp{lstperf}).
    Left panels are for zenith distance range of $<30^\circ$, while the right panels for $30-45^\circ$.
    The bottom panel shows the ratio of sensitivity values:  of the joint analysis sensitivity to MAGIC-only (green triangles) and LST-1-only (red crosses), as well as with respect to the MC sensitivity performance (orange squares).
    E.g. the green points shows that the 3-telescope system, relative to MAGIC alone, can in average detect sources with a flux $\sim 30$\% lower.
    For visibility in the bottom panels points are shifted by $\pm2$\% in the X scale. 
    }
    \label{fig:sens_joint}
\end{figure*}

\begin{figure*}[t]
    \centering
    \includegraphics[width=0.49\textwidth]{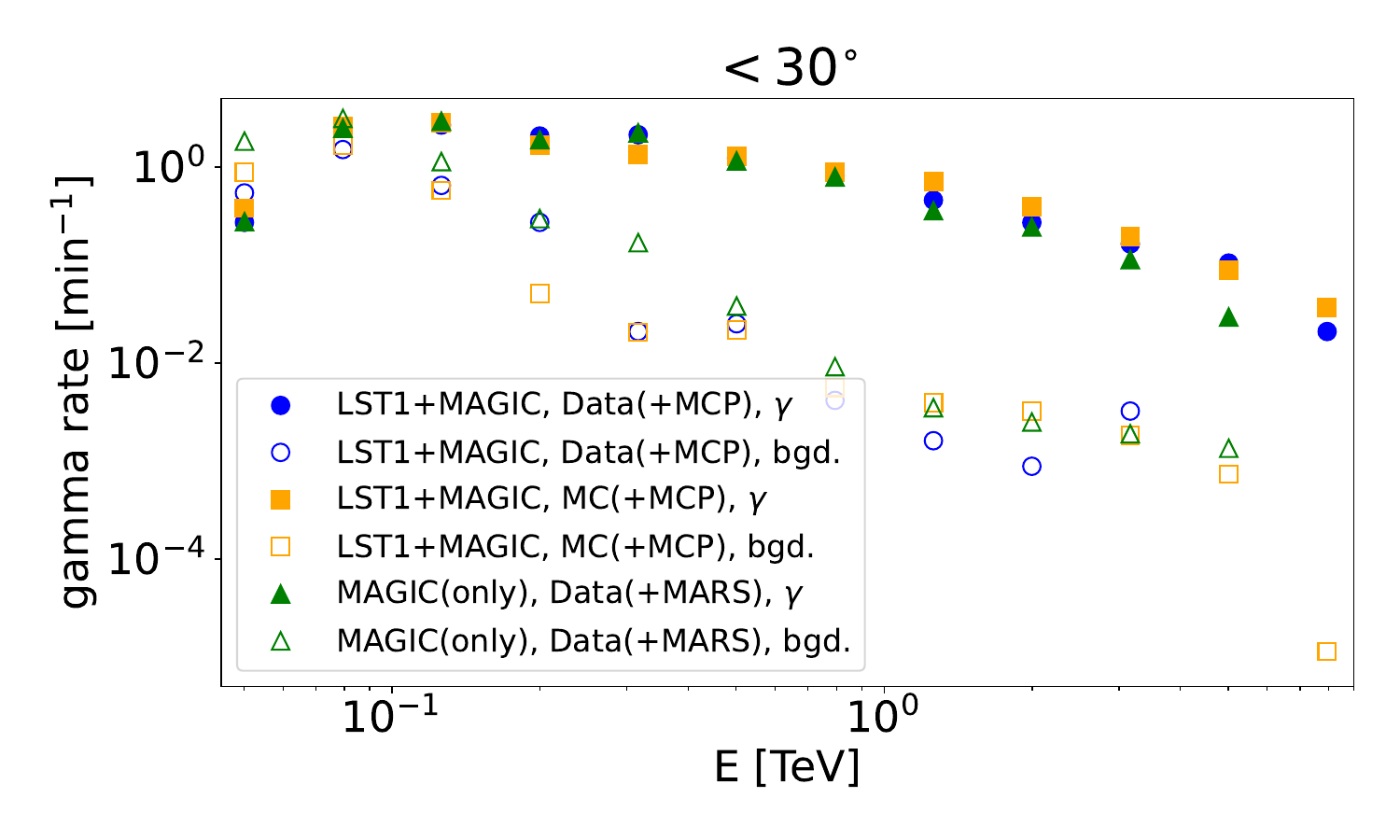}
    \includegraphics[width=0.49\textwidth]{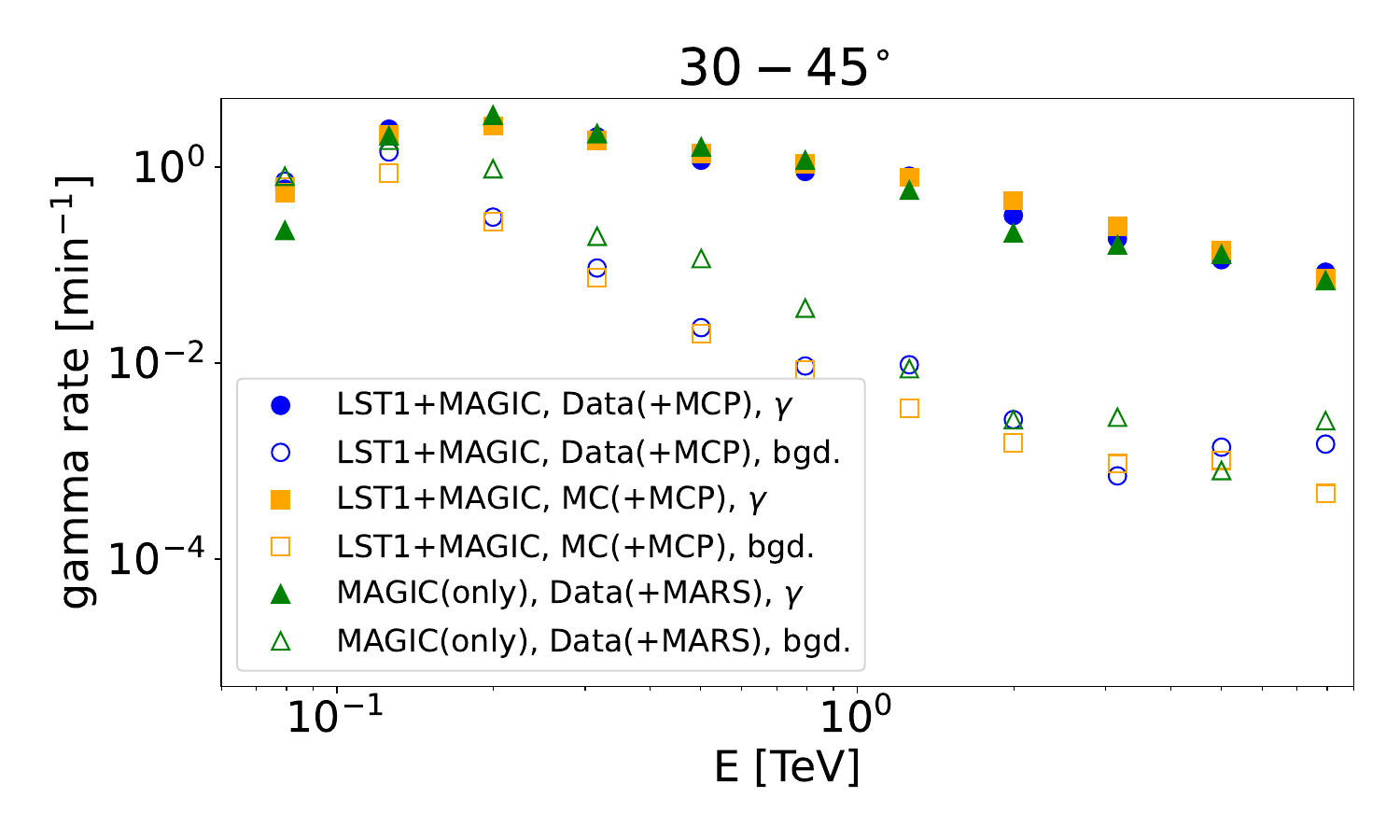}
    \caption{Gamma-ray (excess, full markers) and background (empty markers) rates corresponding to the sensitivity presented in Fig.~\ref{fig:sens_joint}. 
    }
    \label{fig:rates_joint}
\end{figure*}

The resulting sensitivities are compared with the LST-1 standalone and MAGIC stereoscopic sensitivities in Fig.~\ref{fig:sens_joint}.
The corresponding gamma-ray and background rates are presented in Fig.~\ref{fig:rates_joint}.
The MAGIC-only sensitivity curve has been derived from the same dataset as used for the joint analysis. 
As expected, the joint analysis provides significantly better sensitivity.

Using the LST-1+MAGIC joint analysis in the medium energy range (i.e., excluding the first and last two energy bins) allows detection of about 30\% (40\%) weaker fluxes than could be detected with MAGIC-only (LST-1-only) analysis.
This is related to the addition of LST-1 rather than to the different analysis chain as the MAGIC-only performance is compatible with both chains (see~\ref{sec:Monly}).
Such a large gain in performance is expected from the stereoscopic technique when using a small number of telescopes, and is also in line with the previous MC-based study \citep{2019ICRC...36..659D}, which claims 50\% improvement with respect to MAGIC. 
The gain is twofold: first, the addition of the third telescope improves the shower reconstruction, allowing for a more efficient rejection of the background events.
Second, the number of detected gamma-ray events is also increased. 
Since the observations are performed in software-coincidence mode, the trigger-level collection area cannot be increased with the addition of the LST-1.
However, during the regular analysis of data from MAGIC, a fraction of the images do not survive the quality cuts (small showers producing $<50$\,p.e. are typically rejected). 
In the MAGIC-only analysis the rejection of either M1 or M2 image is equivalent to the rejection of the whole event, since it is not possible to perform stereoscopic reconstruction with only one telescope.
However the LST-1 image makes it possible to recover this kind of events (as LST-1+M1 or LST-1+M2 event). 
On average, about 20\% of the reconstructed gamma events has only one image from either M1 or M2 (see Table~\ref{tab:types}). 

Finally, we also compare the joint analysis performance achieved with the Crab observations with the one obtained with MC simulations. 
In the case of low zenith observations, the sensitivity obtained with data and MC simulations is compatible in the whole energy range.
It should be noted that the data curve is based on only 0.8\,hr of data (all the available Crab Nebula joint observation data taken with zenith distance below $30^\circ$ in the data period described by the used MCs), and hence affected by large statistical uncertainties. 
In the case of medium zenith angle observations (where also the statistical uncertainty of the sensitivity is smaller), a $\sim 30$\% mismatch is seen above a few hundred GeV. 
It might be caused by simplifications used in the MC simulations, or by incompleteness of the background MC sample (e.g., missing events with large impact or angular offset, simplification of higher elements treatment). 
It should also be noted that a similar, $\sim 20\%$ mismatch in sensitivity has been observed as well in the earlier simulations of the MAGIC telescopes (\citealp{2012APh....35..435A}, see also \citealp{2022arXiv221209456A}).

\subsection{Angular resolution}

\begin{figure*}[t]
    \centering
    \includegraphics[width=0.49\textwidth]{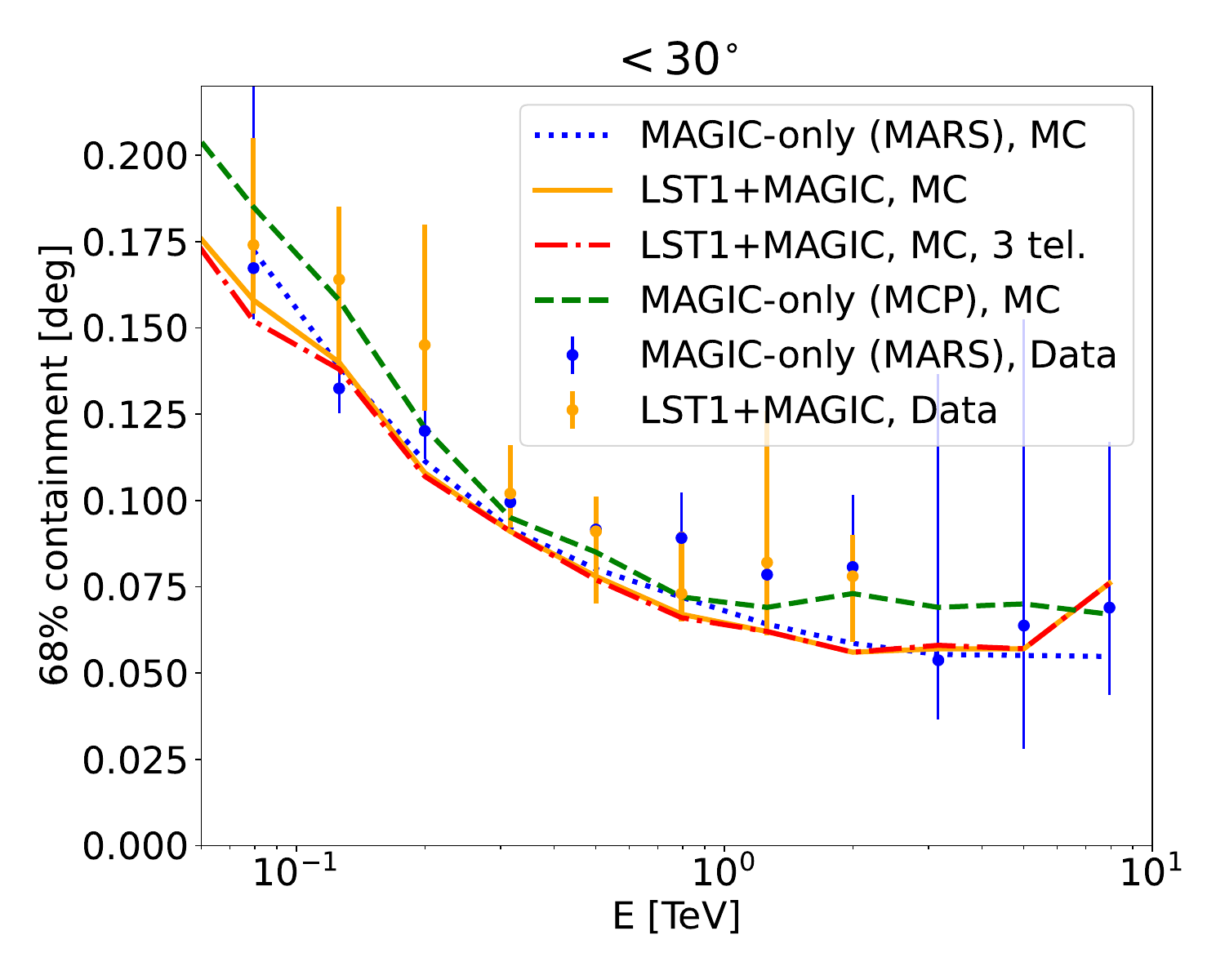}
    \includegraphics[width=0.49\textwidth]{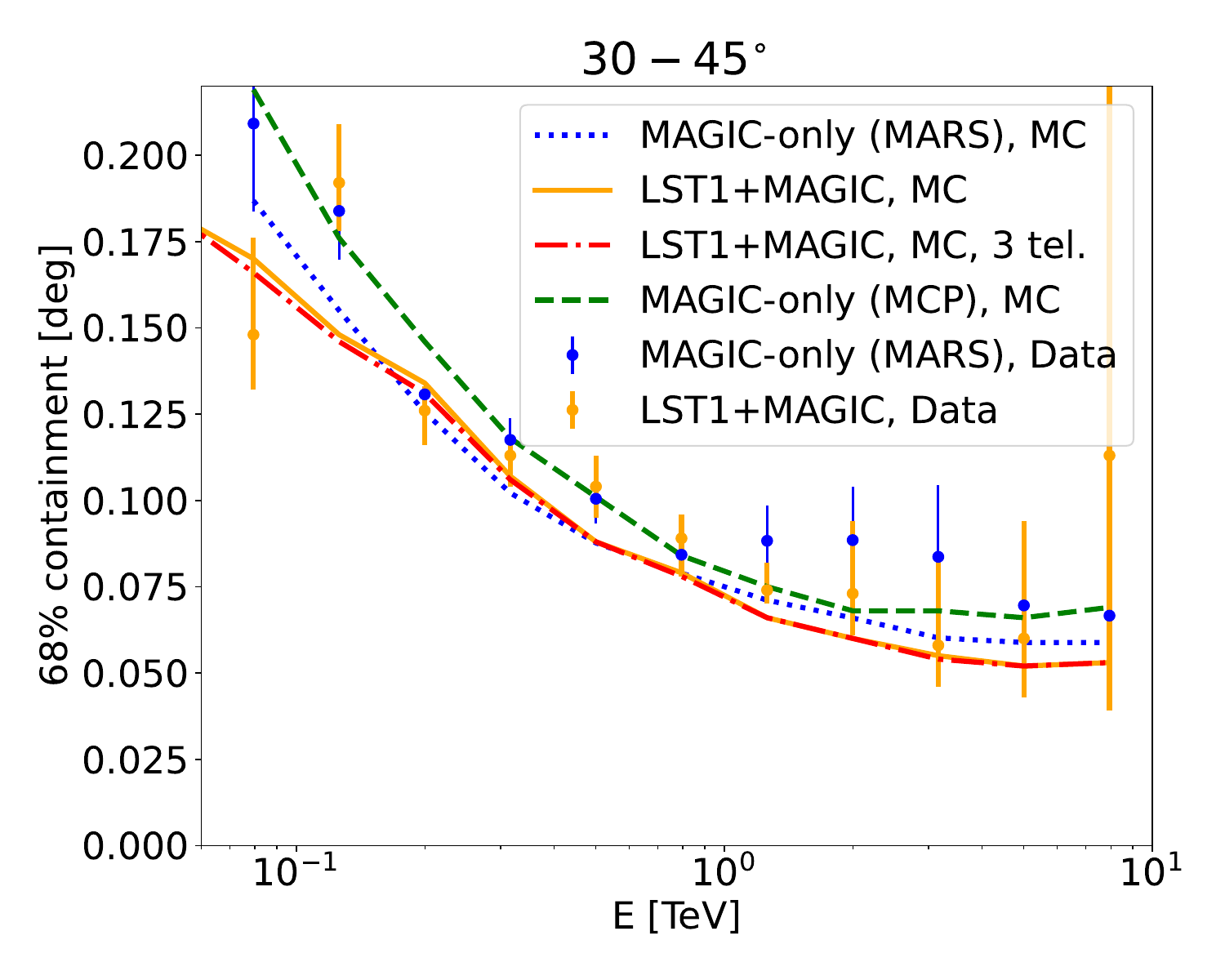}
    \caption{Comparison of the angular resolution (defined as 68\% containment of gamma rays) as a function of the estimated energy of the joint LST-1+MAGIC analysis (orange) from the Crab observations (points) and MC simulations (line) compared to MAGIC-only analysis using MARS (blue) or using MCP (green).
    Joint analysis in which only all-three telescope events are kept is shown in red (in most of the energy range the curve merges with the orange one). 
    Left panel for observations at low zenith angle, right panel for medium zenith angle distance.
    }
    \label{fig:angres_joint}
\end{figure*}

To evaluate the angular resolution of the joint LST-1+MAGIC analysis in every energy bin, the gamma-ray excess has been computed 
as a function of the angular separation to the nominal source position.
To evaluate it in typical circumstances, we apply a 90\% efficiency cut in \textit{gammaness} (the same cut as used for the derivation of the spectral energy distribution of the source).
We follow the commonly used definition of the angular resolution as the angular distance from the source that corresponds to 68\% containment of the point spread function (see Fig.~\ref{fig:angres_joint}). 
To facilitate data/MC comparisons, we use reconstructed energy in both cases, and assume that at $0.4^\circ$ the containment is already 100\% (MC simulations show that the actual containment at $0.4^\circ$ is 96\% in the lowest plotted energy bin, 63 -- 100~GeV). 

For the MAGIC-only analysis the angular resolution is slightly higher ($\sim 10$\%) in the MCP chain as compared to the dedicated MARS analysis. 
It should be noted that the MARS analysis software is optimized for two-telescope observations, and employs slightly different shower reconstruction techniques than \texttt{ctapipe}. 
As a result, at medium and high energies, the angular resolution obtained in the joint analysis with MCP is still similar to the one from MAGIC-alone observations and MARS chain.
However, when comparing instead the joint analysis to MAGIC-only analysis with the same chain, a slightly improvement is seen.  
At the lowest energies the joint analysis reaches a $\sim 10\%$ improvement even with respect to the MARS analysis of MAGIC-only data (only visible in MC curves, as the statistics of the Crab Nebula sample are insufficient for precise evaluation in this energy range). 
This is likely due to the large fraction of dim images at those energies, which are better reconstructed with LST-1 than with MAGIC due to the higher light yield.

We also investigate if further improvements can be achieved by selecting only the events in which all three images are present.
A comparison of the red and orange curves in Fig.~\ref{fig:angres_joint} indicates that the improvement is negligible. 

\subsection{Energy cross-calibration}

\begin{figure}[t]
    \centering
    \includegraphics[width=0.49\textwidth]{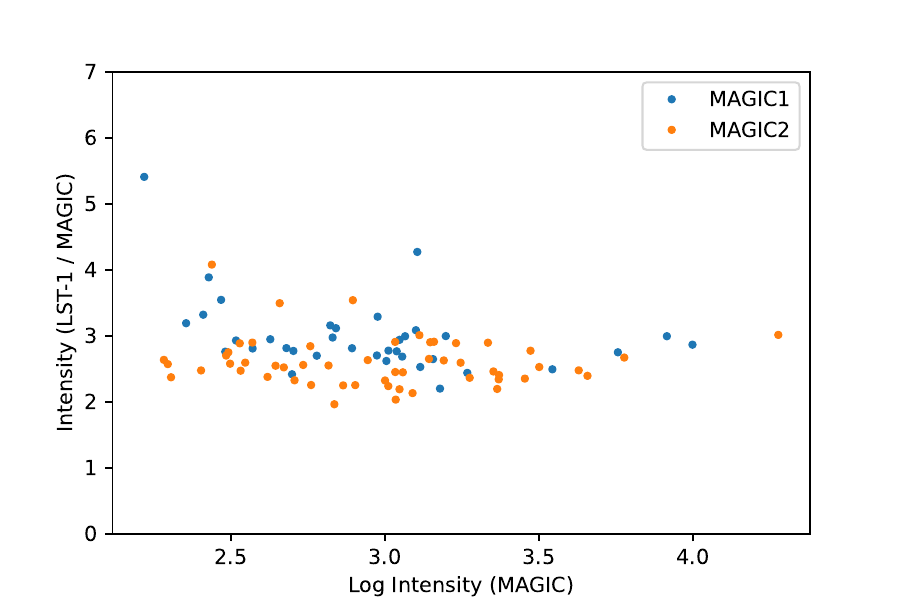}
    \caption{Ratio of the reconstructed image intensities of LST-1 and MAGIC telescopes (blue: MAGIC1, orange: MAGIC2).
    Each point represents a single event. 
    Only events with gammaness above 0.8 and angular distance to the Crab of below $0.1^\circ$, and reconstructed impact parameter to both telescopes within 10\,m are used.
    Only data with zenith distance angle below $45^\circ$, reconstructed impacts below 150~m and reconstructed energy above 300~GeV are used.}
    \label{fig:size_ratio}
\end{figure}

The absolute energy scale calibration is one of the main problems affecting the observations with IACTs.
While current IACTs claim the systematic uncertainty on the energy scale of $\sim 15\%$ \citep{2016APh....72...76A}, for the next generation LST-1 a calibration down to 4\% is possible, but it requires taking into account $\sim 20$ individual systematic effects \citep{2019ApJS..243...11G}.
The small distance between the LST-1 and MAGIC telescopes allows both instruments to see the same showers and to perform joint analysis, but it can also be used to compare the light yield of the telescopes (see \citealp{ho03}).
By selecting events seen at similar impact parameters by LST-1 and one of the MAGIC telescopes, we can compare the light yield of both instruments. 
In Fig.~\ref{fig:size_ratio} we perform such a comparison after applying angular and \textit{gammaness} cuts to select gamma-ray events. 
%
The ratio of the total intensity of the LST-1 images to that of MAGIC-I (MAGIC-II) is 2.99 (2.60), with a standard deviation of 0.57 (0.37).
This is in a rough agreement with the expectations from the larger mirror area and higher photodetector QE (see Table~\ref{tab:pars}).

\begin{figure*}
    \includegraphics[width=0.49\textwidth]{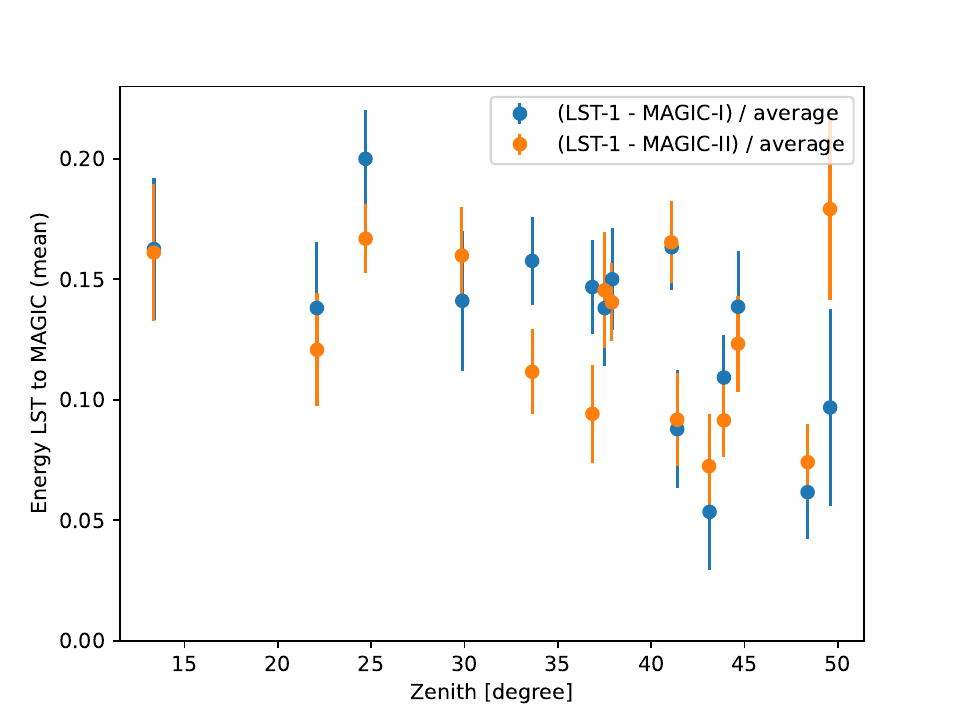}
    \includegraphics[width=0.49\textwidth]{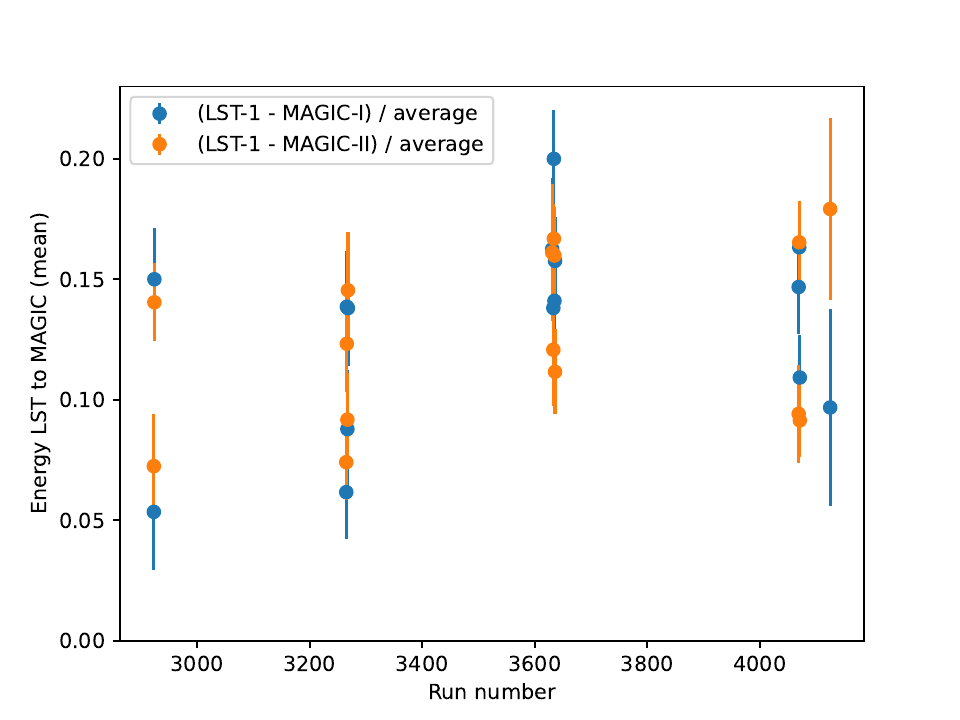}
    \caption{
    Relative difference of estimated energy from LST-1 and MAGIC (blue for MAGIC-I, orange for MAGIC-II) camera images of the same event.
    The energy difference is normalized to weighted average energy obtained from all the telescopes (see Section~\ref{sec:data_analysis}).
    A 90\% \textit{gammaness} efficiency cut, $0.2^\circ$ angular cut between the nominal and reconstructed source position and 0.3 -- 3 TeV estimated energy cut are used. 
    Each point corresponds to average from one run (with the error of the mean reported as error bar). 
    Dependence on the zenith angle of the observation is shown in the left panel, and on the LST-1 run number (increasing with time) is shown in the right panel.
    }
    \label{fig:miscal}
\end{figure*}

In order to investigate more accurately the possible relative miscalibration of the two instruments, we apply a procedure similar to \cite{2016APh....72...76A}.
Namely, we compare the energy estimated using MAGIC camera image parameters to the one estimated for LST-1 using gamma-ray excess events. 
In both cases, stereoscopic parameters (impact and height of the shower maximum) are used as well. 
To avoid any bias present close to the energy threshold, we only use events in the estimated energy range of 0.3 -- 3~TeV, where the energy resolution is almost constant and energy bias is negligible. 
We investigate the relative miscalibration of the two instruments as a function of time or zenith distance of the observations (see Fig.~\ref{fig:miscal}).
The obtained difference in calibration of both instruments is between $\sim 6 - 16 \%$, comparable to the systematic uncertainty on the energy scale of MAGIC.
LST-1 estimates a higher energy than MAGIC, suggesting either an underestimation of the light collection efficiency of LST-1 in MC simulations or an overestimation in the case of MAGIC.
No clear evolution in time is seen, however a hint of zenith dependence (with the miscalibration decreasing at medium zenith) is seen. 
Indeed a constant fit to LST-1 to MAGIC-I (II) zenith dependence provides $\chi^2/N_{\rm dof}=48.8/14$ ($\chi^2/N_{\rm dof}=49.5/14$), while a simple linear fit improves the $\chi^2/N_{\rm dof}$ values to $27.7/13$ ($33.0/13$).

\subsection{Energy resolution}

\begin{figure}[t]
    \centering
    \includegraphics[width=0.49\textwidth]{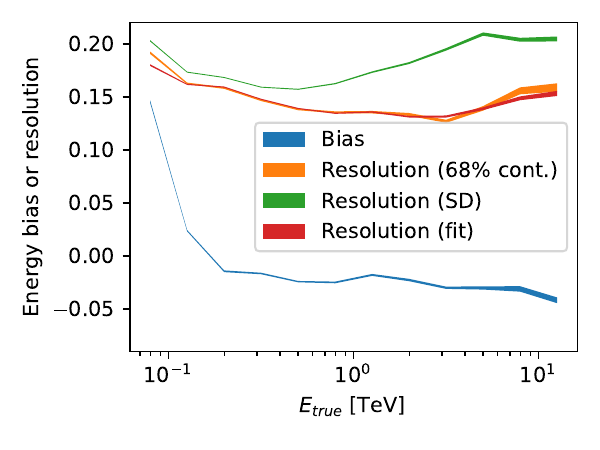}
    \caption{Energy resolution obtained with MC simulations for gamma-ray showers at zenith angle distance of $23.6^\circ$.
    The resolution is calculated as an average 68\% quantile (orange). 
    Standard deviation of the energy misreconstruction is shown in green. 
    The red line is the resolution value obtained from a tail-less (performed in the range mean $\pm$ 2 standard deviations) Gaussian fit.
    The bias of the energy estimation ($(E_{est}-E_{true})/E_{true}$) is shown in blue.
    The thickness of the lines represents the statistical uncertainty. 
    \textit{gammaness} efficiency cut of 90\%, angular distance of $<0.2^\circ$ and \textit{intensity} $>50$~p.e. cuts are applied. }
    \label{fig:en_res}
\end{figure}

In the absence of an external calibrator, the energy resolution of an IACT can only be derived using MC simulations.
To evaluate the performance of the energy reconstruction, we divide the data into bins of true energy, $E_{\rm true}$, and in each bin determine the distribution of the energy dispersion, 
$(E_{\rm est} - E_{\rm true})/E_{\rm true}$.
We define the energy bias as the median of this distribution (we confirmed that using the mean instead would not affect the estimation considerably). 
Similarly to the angular resolution, we define the energy resolution as the 68\% containment area of the dispersion.
We first compute the difference between the median of the distribution and the 16\% and 84\% quantiles, which correspond to the \emph{underestimation} and \emph{overestimation} of the energy estimation.
Next, we compute the 68\% quantile energy resolution as the average (weighted with inversed variance) of these \emph{underestimation} and \emph{overestimation} components. 
For comparison we also compute the energy resolution as the standard deviation of a Gaussian fit to the $(E_{\rm est} - E_{\rm true})/E_{\rm true}$ distribution excluding the tails.
The results are shown in Fig.~\ref{fig:en_res} and summarized in Table~\ref{tab:en_res}.

The energy estimation has nearly no bias down to $\sim 150$~GeV. 
At the lowest energies, the energy resolution is improved with respect to MAGIC-alone observations (cf. \citealp{2022arXiv221203592I}).
The energy resolution in the medium energy range is $\sim 14\%$.
At multi-TeV energies, the energy resolution starts to worsen. 
The 68\% containment definition provides very similar results as a narrow fit definition. 
As the standard deviation is increased by outliers, this definition reports worse energy estimation, in particular at the highest energies.
It should be noted that most of the events above a few TeV in the joint analysis are not fully contained in the camera (i.e., they have pixels surviving the image cleaning in the outermost ring of the camera). 
We also note that the performance of the energy estimation at the highest energies is strongly affected by the details of the training. 
In particular it varies whether the training is done on diffuse gamma rays, or with gamma rays produced in a ring with radius equal to the expected offset used during observations of point-like sources. 
It also depends on the event statistics used in the training. 

\begin{figure}
    \centering
    \includegraphics[width=0.49\textwidth]{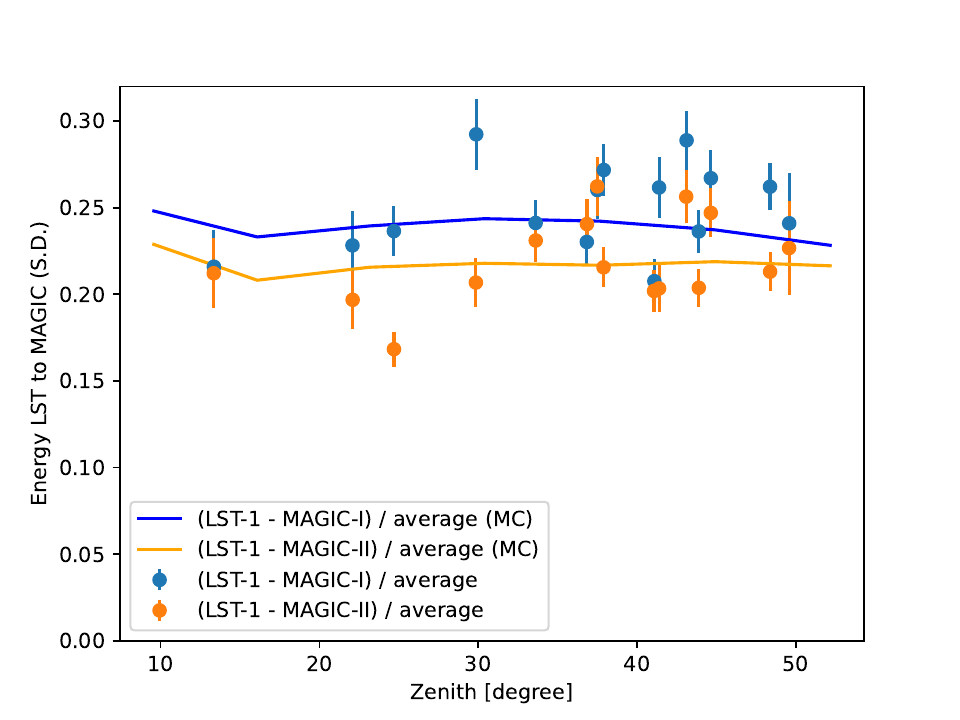}
    \caption{Standard deviation of the relative difference in estimated energy between one of the MAGIC telescopes and LST-1 for the same event.
    Data, MC and quality selection cuts as in Fig.~\ref{fig:miscal}}
    \label{fig:miscal_std}
\end{figure}

In order to validate the accuracy of the claimed energy estimation we perform a similar test as \citet{2016APh....72...76A} (see also Fig.~\ref{fig:miscal}), comparing the spread of the energy estimation between the telescopes. 
Contrary to the energy resolution, such a spread can be then compared with the one obtained from the MC simulations.
As can be seen in Fig.~\ref{fig:miscal_std} the inter-telescope spread of the energy estimation is relatively well reproduced for all the zenith angles, further supporting the reported energy resolution values.  

\subsection{Comparison to single instrument analysis}
The systematic errors affecting measurements performed with IACTs stem both from the complex hardware and from the imperfect knowledge of the atmosphere status, which is considered as part of the detector. 
The systematic uncertainties of MAGIC are divided into three categories: 
uncertainty of the energy scale of $15\%$, pure flux normalization uncertainty of $11-18$\% and uncertainty of the spectral slope of $\pm0.15$ for an assumed power-law spectrum \citep{2016APh....72...76A}. 
In the case of LST-1 standalone observations the uncertainty of the background estimation plays an important role at the lowest energies \citep{lstperf}, because single-telescope observations are characterized by much higher background rates (note that no detailed study of the systematic uncertainties of LST-1 has been performed yet).

In the joint analysis we combine the data from two different IACT systems. 
Such a combination might on one hand increase the systematic uncertainties, in particular due to simplifications needed to merge the simulation and analysis chains for both instruments. 
On the other hand the resulting systematic uncertainties might also decrease because of averaging. 
Moreover, the atmospheric transmission is one of the main sources of systematic uncertainty for IACTs \citep{2012APh....35..435A}, and it can also vary on different time scales.
Due to the physical proximity of MAGIC and LST-1, both instruments share the same contribution from the atmospheric transmission uncertainty. 

While some energy miscalibration between LST-1 and MAGIC has been observed, it is within the claimed systematic uncertainties of MAGIC. 
Moreover, the fact that the reconstructed spectrum agrees closely (within $\sim 10\%$) with the MAGIC and LST-1 standalone results further supports that the systematic uncertainties of the joint analysis are not larger than those of individual instruments. 
The light curve stability analysis requires 12.7\% (7.9\%) relative systematic uncertainties to be consistent with a constant flux on run (day) time scales.
Those numbers are remarkably similar to the ones obtained from such a study for MAGIC-only data: 11\% on run-by-run \citep{2016APh....72...76A} and 7.6\% on day-to-day \citep{2017APh....94...29A}. 
In the case of LST-1-alone observations, the required variable systematic uncertainties on day-to-day time scale are even slightly lower: 6-7\% \citep{lstperf}, however that study was limited only to low zenith distance observations.
Therefore, we conclude that  the systematic uncertainties of the joint analysis are similar to those of MAGIC. 

\section{Conclusions}\label{sec:conc}

We have introduced a new pipeline enabling the joint analysis of LST-1 and MAGIC data.
The LST Collaboration is aiming at performing in the coming years 50\% of the observations together with MAGIC. 
The chain provides common stereoscopic analysis of images corresponding to the same physical shower.
This study is also a path-finder for a stereoscopic analysis chain to be applied to the future CTAO arrays.
As the events are matched by arrival time from the individual LST-1 and MAGIC observations, the rate of events cannot be increased at the trigger level. 
However, the addition of the LST-1 allows us to reconstruct 20\% more events in which one of the MAGIC images is rejected during reconstruction and thus no stereoscopic analysis is possible with MAGIC only.
The analysis provides an improvement of the energy threshold at the reconstruction level by $\sim 15$\% with respect to MAGIC-only observations. 
As a result of such recovered events, and of the improved background rejection for events seen by all three telescopes, the performance of joint observations is greatly improved. Namely, the minimum detectable flux is 30\% (40\%) lower than MAGIC-alone (LST-1-alone) analysis. 
Since in the medium energy range the sensitivity is inversely proportional to the square root of observation time, this corresponds to 2-fold (nearly 3-fold) shortening of the required observation time to reach the same performance of MAGIC-only (LST-1-only) observations. 
Therefore the two systems work more efficiently together than individually, and joint observations are highly valued. 

The other performance parameters, in particular the angular and energy resolution are not strongly affected by the addition of the LST-1 telescope, only minor improvements are seen at the lowest energies. 
The presented analysis is designed to improve the collection area of MAGIC-alone observations by allowing events that would not survive the standard analysis chain of MARS due to one of the images not surviving the quality cuts. 
The additional reconstruction of such worse quality events partially contributes to this lack of significant improvement. 
It is also possible that the peculiar geometrical placement of the three telescopes (obtuse-angled triangle) worsens the efficiency of the shower reconstruction. 
Moreover the comparisons are performed with the MARS analysis of MAGIC data, that is optimized for two-telescope case, and cannot be fully scaled to multiple telescopes.
Further optimization of the analysis for improved angular or energy resolution is possible, possibly at the price of decreased collection area. 

We have performed comparison checks, both against the MAGIC standard simulation software as well as comparisons of the simulation results to the data and showed a rather good agreement. 

The Crab Nebula spectrum obtained from the joint analysis can be described as
$dN/dE=(3.48\pm0.09_{\rm stat})\times 10^{-11} (E/\mathrm{1 TeV})^{-2.49\pm0.03_{\rm stat} - (0.117\pm0.017_{\rm stat}) \ln (E/\mathrm{1 TeV})}
\mathrm{ cm^{-2}s^{-1} TeV^{-1}}
$. 
It is within $\sim 10$\% of the earlier measurements. 
Also the derived flux stability is comparable to the one of MAGIC, pointing to similar systematic uncertainty of the joint analysis. 







\section*{Acknowledgements}

We gratefully acknowledge financial support from the following agencies and organisations:

\bigskip

Ministry of Education, Youth and Sports, MEYS  LM2015046, LM2018105, LTT17006, EU/MEYS CZ.02.1.01/0.0/0.0/16\_013/0001403, CZ.02.1.01/0.0/0.0/18\_046/0016007 and CZ.02.1.01/0.0/0.0/16\_019/0000754, Czech Republic; 
Max Planck Society, German Bundesministerium f{\"u}r Bildung und Forschung (Verbundforschung / ErUM), Deutsche Forschungsgemeinschaft (SFBs 876 and 1491), Germany;
Istituto Nazionale di Astrofisica (INAF), Istituto Nazionale di Fisica Nucleare (INFN), Italian Ministry for University and Research (MUR);
ICRR, University of Tokyo, JSPS, MEXT, Japan;
JST SPRING - JPMJSP2108;
Narodowe Centrum Nauki, grant number 2019/34/E/ST9/00224, Poland;
The Spanish groups acknowledge the Spanish Ministry of Science and Innovation and the Spanish Research State Agency (AEI) through the government budget lines PGE2021/28.06.000X.411.01, PGE2022/28.06.000X.411.01 and PGE2022/28.06.000X.711.04, and grants PGC2018-095512-B-I00, PID2019-104114RB-C31, PID2019-107847RB-C44, PID2019-104114RB-C32, PID2019-105510GB-C31, PID2019-104114RB-C33, PID2019-107847RB-C41, PID2019-107847RB-C43, PID2019-107988GB-C22;
the “Centro de Excelencia Severo Ochoa” program through grants no. CEX2021-001131-S, CEX2019-000920-S;
the “Unidad de Excelencia Mar{\'i}a de Maeztu” program through grants no. CEX2019-000918-M, CEX2020-001058-M;
the “Juan de la Cierva-Incorporaci{\'o}n” program through grant no. IJC2019-040315-I. They also acknowledge the “Programa Operativo” FEDER 2014-2020, Consejer{\'i}a de Econom{\'i}a y Conocimiento de la Junta de Andaluc{\'i}a (Ref. 1257737), PAIDI 2020 (Ref. P18-FR-1580) and Universidad de Ja{\'e}n;
“Programa Operativo de Crecimiento Inteligente” FEDER 2014-2020 (Ref.~ESFRI-2017-IAC-12), Ministerio de Ciencia e Innovaci{\'o}n, 15\% co-financed by Consejer{\'i}a de Econom{\'i}a, Industria, Comercio y Conocimiento del Gobierno de Canarias;
the “CERCA” program of the Generalitat de Catalunya;
and the European Union’s “Horizon 2020” GA:824064 and NextGenerationEU;
We acknowledge the Ramon y Cajal program through grant RYC-2020-028639-I and RYC-2017-22665;
State Secretariat for Education, Research and Innovation (SERI) and Swiss National Science Foundation (SNSF), Switzerland;
The research leading to these results has received funding from the European Union's Seventh Framework Programme (FP7/2007-2013) under grant agreements No~262053 and No~317446.
This project is receiving funding from the European Union's Horizon 2020 research and innovation programs under agreement No~676134.
ESCAPE - The European Science Cluster of Astronomy \& Particle Physics ESFRI Research Infrastructures has received funding from the European Union’s Horizon 2020 research and innovation programme under Grant Agreement no. 824064.

We would like to thank the Instituto de Astrof\'{\i}sica de Canarias for the excellent working conditions at the Observatorio del Roque de los Muchachos in La Palma. The financial support of the German BMBF, MPG and HGF; the Italian INFN and INAF; the Swiss National Fund SNF; the grants PID2019-104114RB-C31, PID2019-104114RB-C32, PID2019-104114RB-C33, PID2019-105510GB-C31, PID2019-107847RB-C41, PID2019-107847RB-C42, PID2019-107847RB-C44, PID2019-107988GB-C22 funded by the Spanish MCIN/AEI/ 10.13039/501100011033; the Indian Department of Atomic Energy; the Japanese ICRR, the University of Tokyo, JSPS, and MEXT; the Bulgarian Ministry of Education and Science, National RI Roadmap Project DO1-400/18.12.2020 and the Academy of Finland grant nr. 320045 is gratefully acknowledged. This work was also been supported by Centros de Excelencia ``Severo Ochoa'' y Unidades ``Mar\'{\i}a de Maeztu'' program of the Spanish MCIN/AEI/ 10.13039/501100011033 (SEV-2016-0588, CEX2019-000920-S, CEX2019-000918-M, CEX2021-001131-S, MDM-2015-0509-18-2) and by the CERCA institution of the Generalitat de Catalunya; by the Croatian Science Foundation (HrZZ) Project IP-2016-06-9782 and the University of Rijeka Project uniri-prirod-18-48; by the Deutsche Forschungsgemeinschaft (SFB1491 and SFB876); the Polish Ministry Of Education and Science grant No. 2021/WK/08; and by the Brazilian MCTIC, CNPq and FAPERJ.

This work was conducted in the context of the CTA LST Project. 
%
%
A.~Berti: development and maintenance of the module \texttt{ctapipe\_io\_magic}, implementation of processing of MAGIC calibrated data within \texttt{magic-cta-pipe}, general maintenance of the \texttt{magic-cta-pipe} package;
F.~Di Pierro: development of the simultaneous observations' database and of the first versions of \texttt{ctapipe\_io\_magic}, cross-checks with the MARS analysis, coordination of the analysis cross-checks;
E.~Jobst: implementation of tests for low-level comparison of shower images and parameters for MAGIC-only analysis, upgrade of \texttt{ctapipe\_io\_magic} to new releases of \texttt{ctapipe}, test of \texttt{magic-cta-pipe} for MAGIC-only analysis;
Y.~Ohtani: responsible for most of \texttt{magic-cta-pipe} analysis code, many tests and optimization of the pipeline, calculation of performance parameters;
J.~Sitarek: reoptimization and processing of Monte Carlo simulations, calculation of performance parameters, paper drafting; 
Y.~Suda: contribution to the first iteration of implementation of MAGIC to \texttt{sim\_telarray} and coordinating the group from the MAGIC side together with F.~Di Pierro;
E.~Visentin: performance parameters cross-check analysis.
The rest of the authors have contributed in one or several of the following ways: design, construction,
maintenance and operation of the instrument(s); preparation and/or evaluation of the
observation proposals; data acquisition, processing, calibration and/or reduction;
production of analysis tools and/or related Monte Carlo simulations; discussion and
approval of the contents of the draft.
JS would like to thank ICRR for excellent working conditions during final stages of preparation of this manuscript. 
YS' work was supported by JSPS KAKENHI Grant Number JP21K20368 and JP23K13127.
The authors would like to thank anonymous journal referee for the feedback which helped to improve the paper. 

\appendix

\section{Collection area at different analysis stages}\label{sec:Aeff}

\begin{figure}[t]
    \centering
    \includegraphics[trim = 10 0 25 25, clip,width=0.49\textwidth]{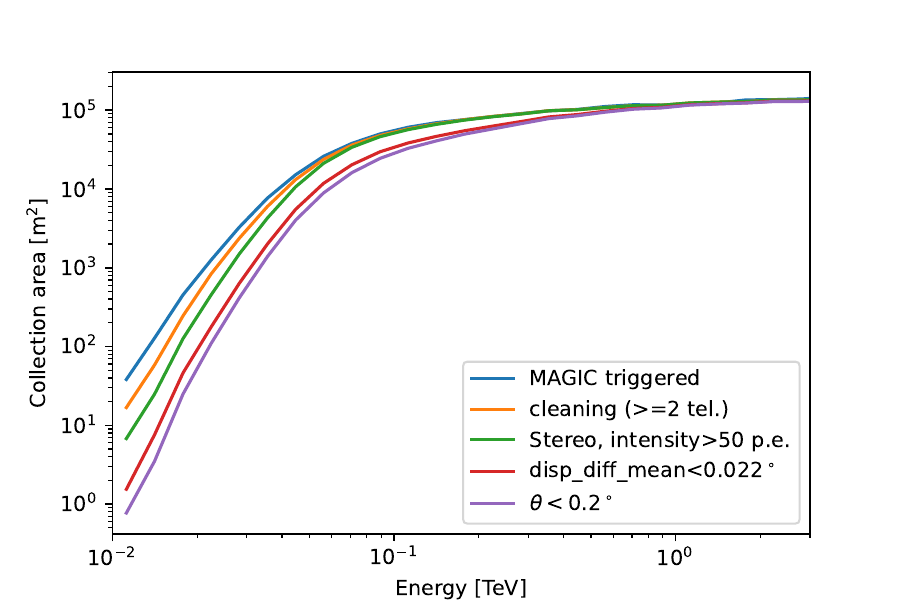}\\
    \includegraphics[trim = 10 0 25 25, clip,width=0.49\textwidth]{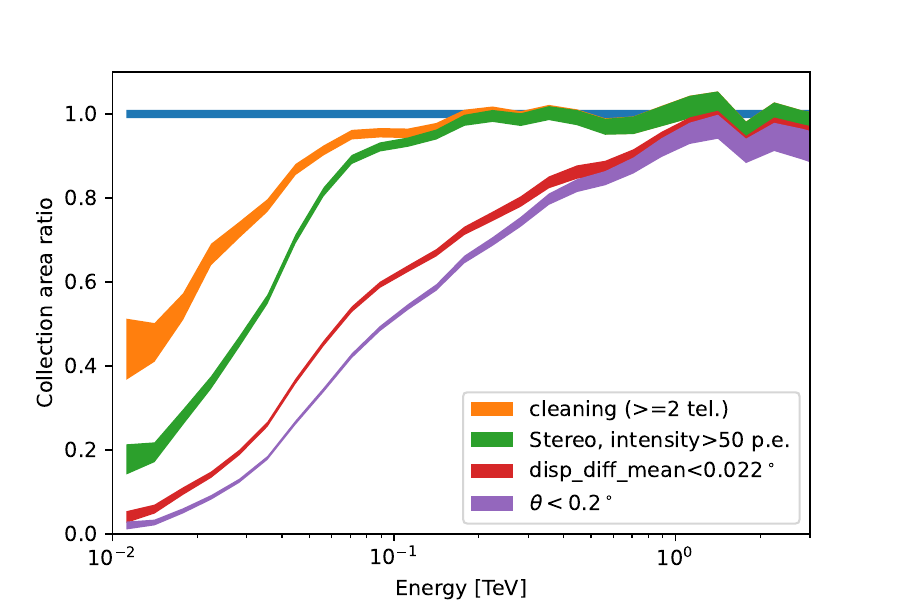}
    \caption{Collection area at zenith angle of 10$^\circ$ for gamma rays at wobble offset of $0.4^\circ$ for different stages of analysis: events triggered by both MAGIC telescopes (blue), 
    events with at least two images surviving cleaning (orange), 
    stereoscopic reconstruction with at least two images with intensity above 50 p.e. (green), 
    cut in consistency of reconstructed arrival direction from different images (red), 
    cut in angular distance between the true and reconstructed shower direction (violet). 
    The bottom plot shows the ratio of the collection area at different stages to the MAGIC one at trigger level, with the band width reporting the statistical uncertainty.
    }
    \label{fig:stages}
\end{figure}

In Fig.~\ref{fig:stages} we present the changes of the energy-dependent collection area at different analysis stages.
Similarly to \citet{lstperf} a large drop of the collection area at the lowest energies ($\lesssim 70$~GeV), and the resulting shift of the energy threshold, occurs due to required image cleaning and \emph{intensity} quality cut. 
The second quality cut in the agreement of reconstructed positions from different telescopes results in further drop of the collection area, visible up to a few hundred GeV. 
Nevertheless, that cut improves considerably angular resolution, such that the effect of the $\theta<0.2^\circ$ cut is much milder. 

\section{Validation of the simulation settings}\label{sec:mccomp}
\begin{figure*}[pt!]
    \centering
    \includegraphics[width=0.99\textwidth]{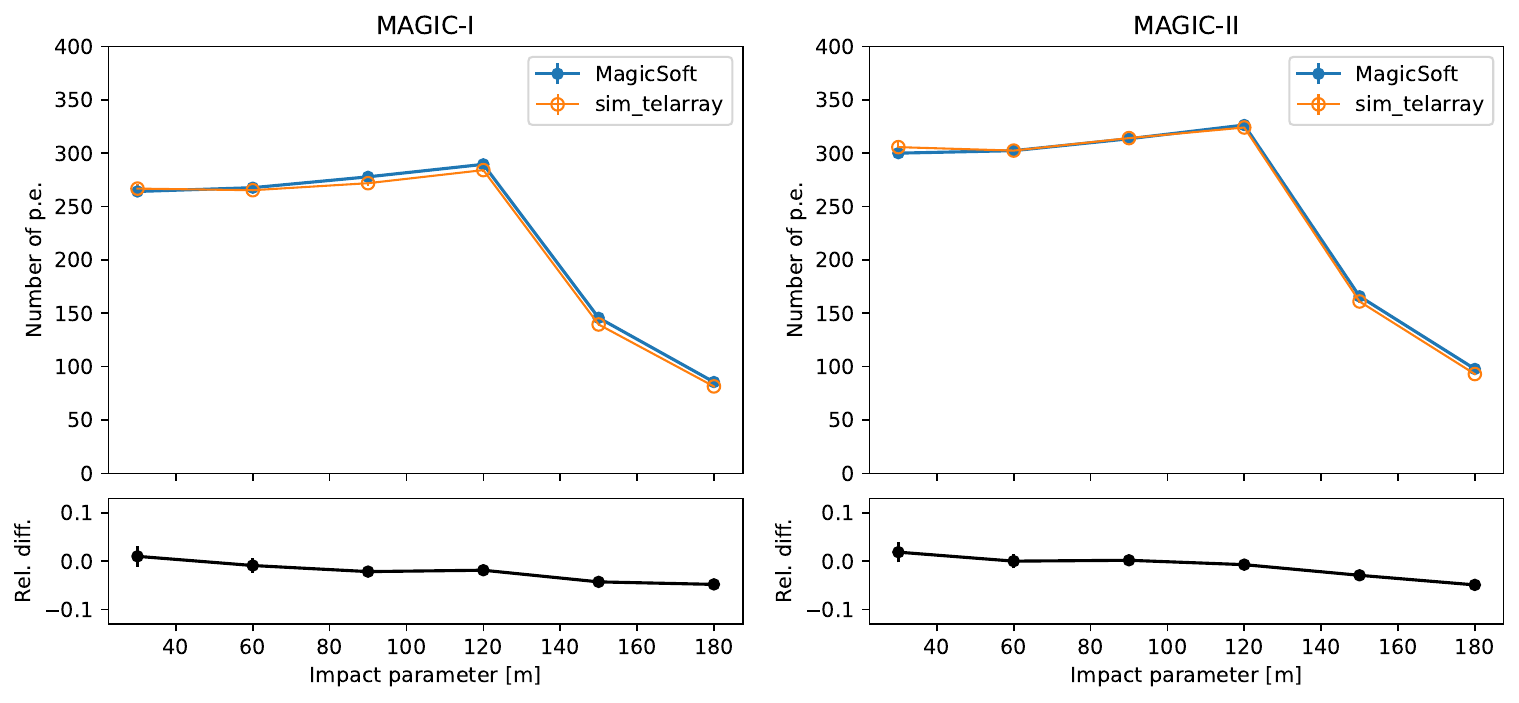}
    \caption{Comparison of the true number of p.e. obtained with \texttt{MagicSoft} (blue) and \texttt{sim\_telarray} (orange) for 100~GeV gamma rays for MAGIC-I (left) and MAGIC-II (right).
    In the bottom panel the relative difference of \texttt{sim\_telarray} with respect to \texttt{MagicSoft} is shown.}
    \label{fig:npe_comp}
\end{figure*}

\begin{figure*}[pt!]
    \centering
    \includegraphics[width=0.99\textwidth]{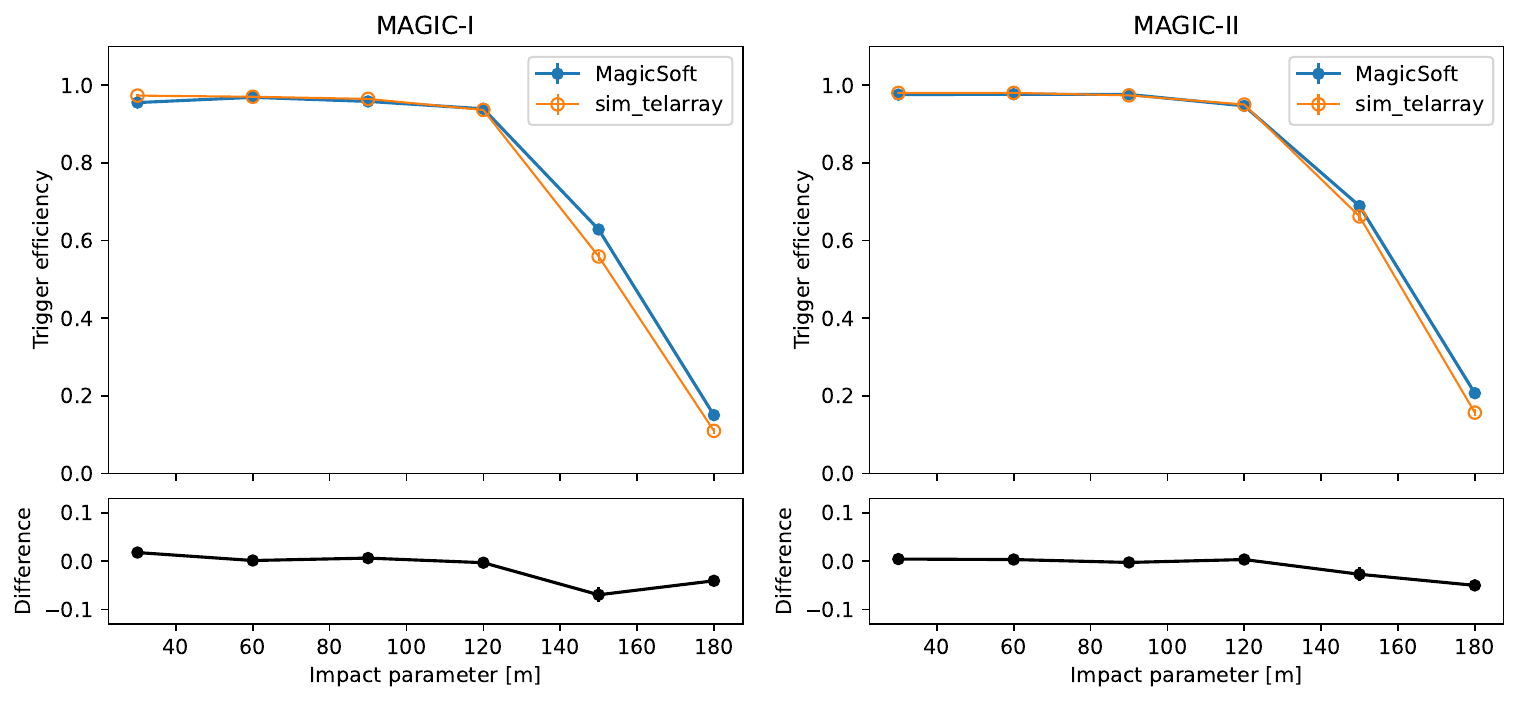}
    \caption{Comparison of the trigger efficiency obtained with \texttt{MagicSoft} (blue) and \texttt{sim\_telarray} (orange) for 100~GeV gamma rays for MAGIC-I (left) and MAGIC-II (right).
    In the bottom panel the difference of \texttt{sim\_telarray} with respect to \texttt{MagicSoft} is shown.}
    \label{fig:trig_comp}
\end{figure*}

To validate the parameter translation procedure and to assure that the necessary simplifications do not significantly affect  the results, a comparison of dedicated MC samples (produced independently with \texttt{MagicSoft} and \texttt{sim\_telarray}) was performed. 
For each program we generated\footnote{Those special simulations were done not using the array geometry as shown in Fig.~\ref{fig:tel_loc}, but a ``virtual'' array of MAGIC telescopes located at the distance, 30m, 60m, ..., 180m from a fixed shower axis impact point. 
For higher impact distances the trigger efficiency drops dramatically. } 2000 vertical gamma-ray showers of energy 100~GeV, at impact parameters uniformly distributed in the range 30 -- 180\,m.
In Fig.~\ref{fig:npe_comp} we present the comparison of true number of p.e. obtained with both chains. 
The two chains are agreeing on the total observed light yield within $\sim2\%$ within the lightpool hump (i.e. for impacts $\lesssim120$~m, and within 5\% in the tail of the shower. 
Similarly good agreement is also achieved in the trigger efficiency (ratio of triggered and simulated events, see Fig.~\ref{fig:trig_comp}).

\section{MAGIC-only performance with MCP}\label{sec:Monly}

\begin{figure*}[pt]
    \centering
    \includegraphics[width=0.99\textwidth]{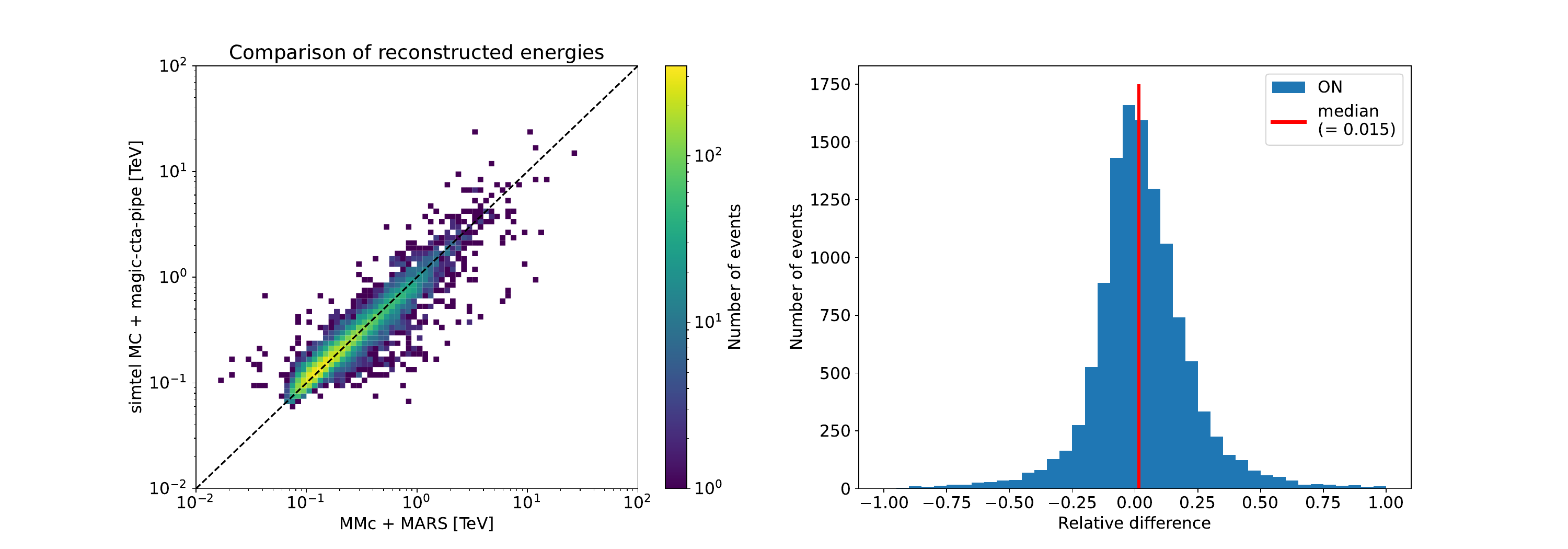}
    \caption{Comparison of the energy estimate of the same MAGIC-only events with MARS and MCP chain. Left panel: energy estimated with MCP chain vs energy estimated with MARS. Right panel: relative difference of MCP-estimated energy with respect to the MARS-estimated one. 
    Only gamma-like events with MARS hadronness value of $<0.2$ and intensity of each image above 100 p.e. are used.}
    \label{fig:mcp_mars_energy}
\end{figure*}

MCP analysis chain can be also applied to MAGIC-only events.
Such a use case has a limited practical application, because the high level MAGIC-only analysis can also be performed in the CTAO-like framework starting from the so-called DL3 data level \citep{2019A&A...625A..10N}, however it turned out to be a useful tool in debugging and comparing the performance of the MAGIC standard chain and MCP. 
To validate the analysis procedures we performed such an analysis of a MAGIC Crab Nebula sample. 
The data are taken on the same nights as the sample used for joint analysis. 
However, because of lack of the simultaneity condition they amount to a larger duration of 6.6 hrs of effective time (out of which 2.2~hrs are taken in zenith range $<30^\circ$ and 3.5~hrs in $30-45^\circ$). 
In Fig.~\ref{fig:mcp_mars_energy} we compare the energy estimation of the same gamma-like events processed with MCP and with the standard MARS analysis chains.  
The MCP chain for MAGIC-only analysis uses the same (\texttt{sim\_telarray}-based) MC simulations as for the joint analysis, however only MAGIC telescopes are selected. 
There is no visible bias between the two analysis chains - the average energy estimate is consistent within $\sim 2$\%. 

\begin{figure*}[t]
    \centering
    \includegraphics[width=0.49\textwidth]{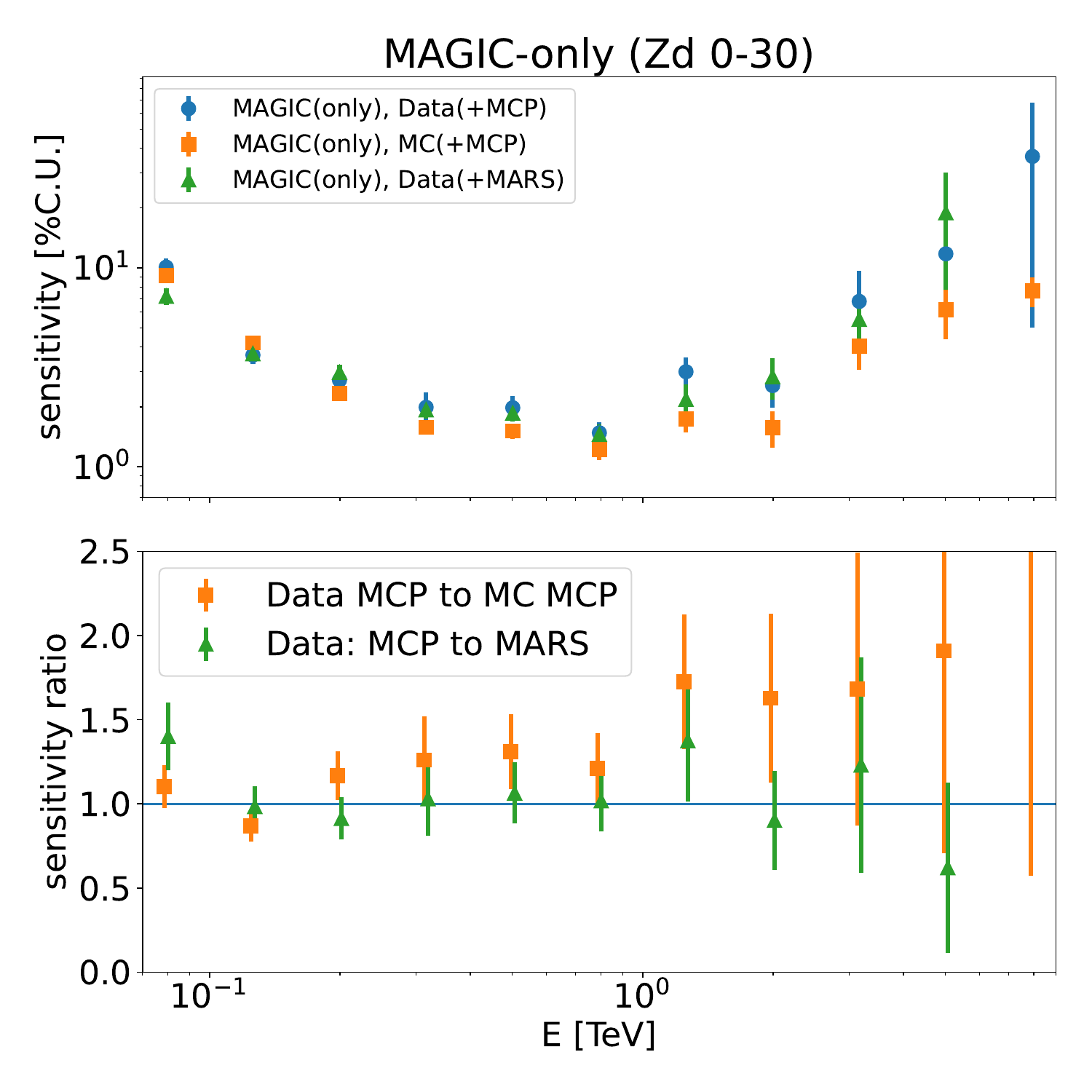}
    \includegraphics[width=0.49\textwidth]{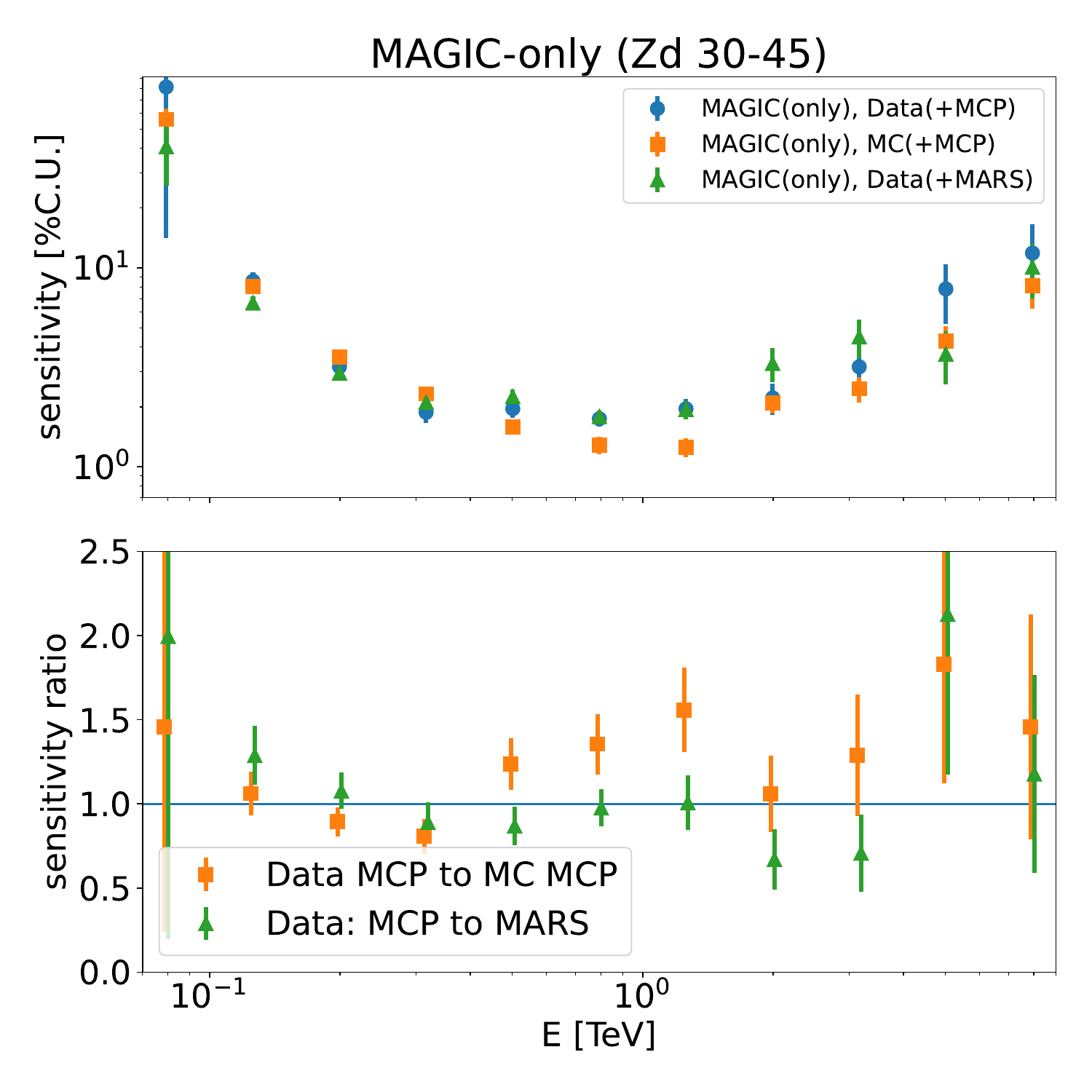}
    \caption{Differential sensitivity of MAGIC-only observations of MCP analysis chain (blue circles for data, orange squares for MC) compared with the standard MARS analysis over the same data sample (green triangles) for low zenith (left) and medium zenith (right) case. 
    The bottom panel shows the sensitivity ratios (for visibility of uncertainty bars, the points are shifted in the X axis by $\pm1\%$).
    }
    \label{fig:sens_Monly}
\end{figure*}

In Fig.~\ref{fig:sens_Monly} we present the differential sensitivity comparison with such a data set. 
In the medium energy range the performance of both chains is similar down to the statistical errors (however a hint of possibly worse performance of MCP is seen at the lowest energies). 
Comparing the sensitivities computed using the data obtained by observations and MC simulations the differences for MAGIC-only analysis are typically $\sim 20-30\%$, except at the highest energies for low-zenith case, where the MC sensitivity uncertainty is very large.
Similar differences at mid energies between the data and MC are also reported in the integral sensitivity of \cite{2012APh....35..435A}. 

\section{Data/MC comparisons with background events}\label{sec:comp_bgd}
\begin{figure*}[tp!]
    \centering
    \includegraphics[width = 0.99\textwidth]{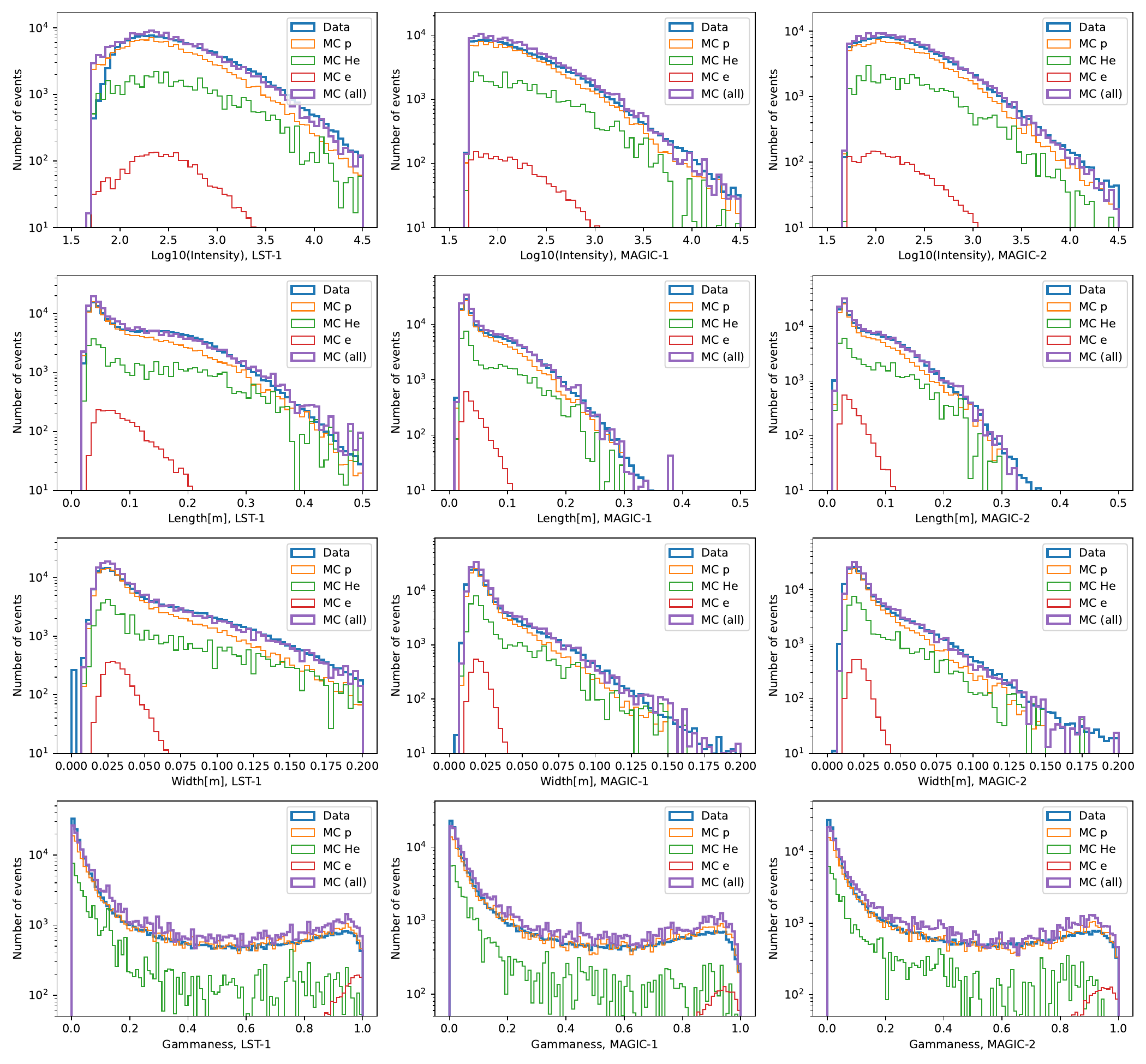}
    \includegraphics[width = 0.99\textwidth]{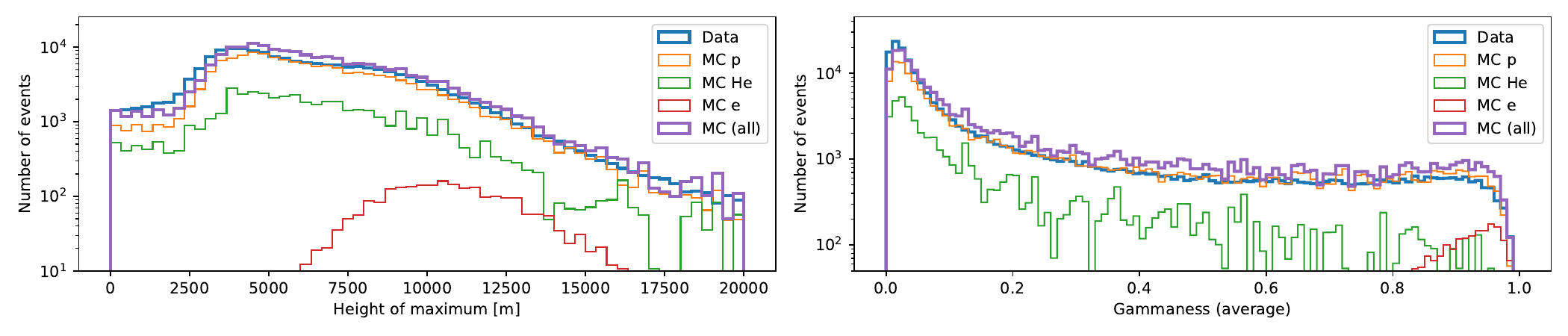}
    \caption{Comparison of image parameters between the Data and MC simulations (for zenith distance below 30$^\circ$).
    Top four rows of panels show \textit{intensity}, \textit{length}, \textit{width} and individual telescope \textit{gammaness} (from top to bottom) for LST-1 (left), MAGIC-I (middle) and MAGIC-II (right).
    The bottom row shows stereoscopic parameters: height of the shower maximum (left) and averaged \textit{gammaness} (right).
    In all the panels thick blue like shows the data, thick magenta line shows the sum of all MC components. 
    Thin lines show the individual components: protons (orange), helium (with an additional correction for heavier elements (green) and all-electrons (red). }
    \label{fig:comp_bgd}
\end{figure*}

Similarly to the comparisons using the gamma-ray excess, we also compare the bulk of the observed events (cosmic ray background) with the MC simulations. 
While such a comparison is less sensitive to the optical telescope parameters it is more direct as it does not require any preselection of events. 
The results of the comparisons are shown in  Fig.~\ref{fig:comp_bgd}. 
The main contribution in the background events before gamma-selection cuts is caused by protons, for which we adopt the spectrum from \cite{2019ICRC...36..163Y}. 
However, we also take into account helium (with a correction for heavier elements, see Section~\ref{sec:sens} for details) and electrons. 
Only events with reconstructed direction within $1^\circ$ from the camera center are used. 
Moreover, to avoid contamination from the Crab Nebula gamma rays, a region with a radius of $0.2^\circ$ around the nominal source position has been excluded. 
We also exclude MAGIC-only events without an LST-1 counterpart. 
The obtained normalization of the distributions is in agreement with the cosmic-ray measurements. 
The used cut of \textit{intensity} $>50$ p.e. is sufficient to reproduce properly the \textit{intensity} distribution of both MAGIC telescopes. 
In the case of LST-1 however a slight mismatch $\lesssim80$ p.e. is visible, which was also reported in \cite{lstperf} and explained as an effect of less stable trigger thresholds in the data until August 2021.
Both \textit{width} and \textit{length} parameter distributions are closely matching between the data and MC simulations. 
The \textit{gammaness} distribution of individual telescopes (as well as the telescopes-averaged values) is relatively well reproduced with MC simulations. 
In the case of the height of the shower maximum, the distribution shape is reproduced well, however a small shift is present as well.

\section{Additional tables}
For convenience and possible comparisons with other instruments in this section we report the numerical values of the performance parameters. 
In tables~\ref{tab:sens_lowzd} and \ref{tab:sens_midzd} we summarize the gamma-ray excess rates, background rates and derived sensitivity from Crab Nebula data sample and MC simulations.
\begin{table*}[p!]
    \centering
    \begin{tabular}{c|c|c|c|c}
    $E_0$ & Gamma rate & Background rate & Sensitivity (data) & Sensitivity (MC) \\
    $\mathrm{[TeV]}$ & (data), $[\mathrm{min^{-1}}]$ &  (data), $[\mathrm{min^{-1}}]$ &  [\%C.U.] & $[\mathrm{10^{-12} cm^{-2} s^{-1} erg}]$ \\\hline
0.0501 & 0.27$\pm$0.17  & 0.54$\pm$0.11 &28.0$\pm$19.0 & 21.1$\pm$ 2.1 \\
0.0794 & 2.59$\pm$0.34  & 1.50$\pm$0.18 &4.81$\pm$0.82 & 4.41$\pm$ 0.22 \\
0.126 & 2.68$\pm$0.29  & 0.65$\pm$0.12 &3.08$\pm$0.51 & 2.41$\pm$ 0.11 \\
0.2 & 2.07$\pm$0.23  & 0.272$\pm$0.075 &2.61$\pm$0.54 & 1.237$\pm$ 0.089 \\
0.316 & 2.13$\pm$0.21  & 0.021$\pm$0.021 &0.77$\pm$0.4 & 0.946$\pm$ 0.084 \\
0.501 & 1.23$\pm$0.16  & 0.025$\pm$0.010 &1.44$\pm$0.36 & 0.882$\pm$ 0.086 \\
0.794 & 0.81$\pm$0.13  & 0.0041$\pm$0.0017 &1.03$\pm$0.27 & 0.610$\pm$ 0.073 \\
1.26 & 0.458$\pm$0.098  & 0.0016$\pm$0.0011 &1.29$\pm$0.54 & 0.537$\pm$ 0.087 \\
2.0 & 0.271$\pm$0.075  & 0.00088$\pm$0.00088 &1.8$\pm$1.0 & 0.69$\pm$ 0.13 \\
3.16 & 0.164$\pm$0.059  & 0.0032$\pm$0.0023 &4.6$\pm$2.4 & 0.82$\pm$ 0.21 \\
5.01 & 0.104$\pm$0.047  & -- &3.2$\pm$1.4 & 0.96$\pm$ 0.32 \\
7.94 & 0.021$\pm$0.021  & -- &16.0$\pm$16.0 & 0.80$\pm$ 0.14 \\
    \end{tabular}
    \caption{Rates and sensitivity values for Crab Nebula observations and MC simulations at zenith distance $<30^\circ$, as plotted in Figures \ref{fig:sens_joint} and \ref{fig:rates_joint} (left panels).\\ 
    \textit{Note:}
    Rates are integrated in 0.2 decades centered on $E_0$ value. 
    Data sensitivities are provided in the percentage of Crab Nebula flux, while MC sensitivities in SED units.}
    \label{tab:sens_lowzd}
\end{table*}
\begin{table*}[p!]
    \centering
    \begin{tabular}{c|c|c|c|c}
    $E_0$ & Gamma rate & Background rate & Sensitivity (data) & Sensitivity (MC) \\
    $\mathrm{[TeV]}$ & (data), $[\mathrm{min^{-1}}]$ &  (data), $[\mathrm{min^{-1}}]$ &  [\%C.U.] & $[\mathrm{10^{-12} cm^{-2} s^{-1} erg}]$ \\\hline
0.0794 & 0.59$\pm$0.12  & 0.715$\pm$0.072 &14.6$\pm$3.4 & 13.0$\pm$ 0.96 \\
0.126 & 2.43$\pm$0.19  & 1.42$\pm$0.10 &4.99$\pm$0.51 & 3.86$\pm$ 0.16 \\
0.2 & 2.65$\pm$0.15  & 0.307$\pm$0.047 &2.16$\pm$0.24 & 1.747$\pm$ 0.072 \\
0.316 & 2.03$\pm$0.13  & 0.093$\pm$0.026 &1.6$\pm$0.26 & 1.206$\pm$ 0.064 \\
0.501 & 1.171$\pm$0.093  & 0.0229$\pm$0.0057 &1.46$\pm$0.22 & 0.798$\pm$ 0.047 \\
0.794 & 0.899$\pm$0.081  & 0.0093$\pm$0.0016 &1.29$\pm$0.16 & 0.595$\pm$ 0.047 \\
1.26 & 0.806$\pm$0.076  & 0.0096$\pm$0.0016 &1.45$\pm$0.19 & 0.458$\pm$ 0.048 \\
2.0 & 0.319$\pm$0.048  & 0.00264$\pm$0.00076 &2.21$\pm$0.46 & 0.458$\pm$ 0.062 \\
3.16 & 0.185$\pm$0.036  & 0.0007$\pm$0.00049 &2.46$\pm$0.99 & 0.525$\pm$ 0.089 \\
5.01 & 0.113$\pm$0.029  & 0.00138$\pm$0.00056 &5.0$\pm$1.6 & 0.67$\pm$ 0.15 \\
7.94 & 0.084$\pm$0.025  & 0.00148$\pm$0.00086 &6.8$\pm$2.9 & 0.70$\pm$ 0.20 \\
    \end{tabular}
    \caption{As in Table~\ref{tab:sens_lowzd} but for zenith distance $30-45^\circ$ (see also Figures \ref{fig:sens_joint} and \ref{fig:rates_joint}, right panels).}
    \label{tab:sens_midzd}
\end{table*}
In Fig.~\ref{tab:en_res} we report the energy resolution derived with different definitions. 

\begin{table*}[p!]
    \centering
    \begin{tabular}{c|c|c|c}
E & Res. 68\% & Res. (S.D.) & Res. (fit) \\ 
 $[$TeV$]$ & $[\%]$ & [\%] & [\%] \\\hline
0.0794 & 19.2$\pm$0.1 &  20.27$\pm$0.06 &  16.16$\pm$0.05 \\
0.126 & 16.26$\pm$0.07 &  17.32$\pm$0.04 &  15.63$\pm$0.04 \\
0.2 & 15.83$\pm$0.08 & 16.82$\pm$0.04 &  15.34$\pm$0.04 \\
0.316 & 14.68$\pm$0.08 & 15.91$\pm$0.04 &  14.5$\pm$0.04 \\
0.501 & 13.82$\pm$0.08 & 15.7$\pm$0.05 &  13.99$\pm$0.04 \\
0.794 & 13.54$\pm$0.09 & 16.24$\pm$0.06 &  13.84$\pm$0.05 \\
1.26 & 13.5$\pm$0.1 & 17.32$\pm$0.07 &  13.87$\pm$0.06 \\
2.0 & 13.3$\pm$0.1 & 18.18$\pm$0.09 &  13.57$\pm$0.07 \\
3.16 & 12.7$\pm$0.1 & 19.5$\pm$0.1 &  13.63$\pm$0.08 \\
5.01 & 13.9$\pm$0.2 & 20.9$\pm$0.2 &  14.4$\pm$0.1 \\
7.94 & 15.6$\pm$0.3 & 20.4$\pm$0.2 &  15.0$\pm$0.1 \\
12.6 & 15.9$\pm$0.4 & 20.4$\pm$0.2 &  15.2$\pm$0.2 \\
    \end{tabular}
    \caption{Energy resolution at zenith distance $23.6^\circ$ corresponding to Fig.~\ref{fig:en_res}.\\
    \textit{Note:}
    The columns report: true energy, 68\% containment resolution, standard deviation (S.D.) and tail-less fit.}
 \label{tab:en_res}
\end{table*}

\end{document}